\documentclass[10pt,a4paper]{article}
\usepackage[T1]{fontenc}
\usepackage[margin=2cm]{geometry}
\usepackage{graphicx,xcolor,array,caption}
\usepackage[authoryear,round]{natbib}
\newcounter{subfigure}[figure]
\usepackage{url,hyperref,microtype}
\hypersetup{hidelinks}

\usepackage{amsmath,amssymb,amsfonts,upref,endnotes}
\usepackage{amsthm}
\usepackage{marginnote}
\usepackage{soul}
\usepackage[figuresright]{rotating}
\usepackage[nolist]{acronym}
\usepackage{multirow}
\usepackage{booktabs}
\usepackage{cleveref}
\usepackage{tikz}
\usepackage{lmodern}
\usepackage{listings}

\usepackage{algorithm}
\usepackage{algorithmicx}
\usepackage{algpseudocode}

\usepackage{glossaries}

\makeatletter
\AtBeginEnvironment{figure}{\setcounter{subfigure}{0}}
\AtBeginEnvironment{figure*}{\setcounter{subfigure}{0}}
\renewcommand{\thesubfigure}{\Alph{subfigure}}
\renewcommand{\p@subfigure}{\number\numexpr\value{figure}+1\relax}

\crefname{subfigure}{Figure}{Figures}
\Crefname{subfigure}{Figure}{Figures}
\newcommand{\subfloat}[2][]{%
  \refstepcounter{subfigure}%
  \begin{tabular}{@{}c@{}}%
    \textbf{(\thesubfigure)}\\[-0.2em]%
    #2%
  \end{tabular}%
}
\makeatother

\newglossaryentry{graph}{
  name=graph,
  description={A collection of vertices and edges. Additionally, in this paper, all graphs are finite, labeled and directed, unless specified otherwise}
}

\newglossaryentry{vertex}{
  name=vertex,
  description={A distinct entity in a graph, which can be connected to other vertices via edges}
}

\newglossaryentry{edge}{
  name=edge,
  description={The link between two vertices, including the connection, the label for identification, and contextual properties in describing this relationship}
}

\newglossaryentry{label}{
  name=label,
  description={As a noun: a symbolic unique identifier of an edge, either semantic or not. --- As a verb: the act of attaching a label to an edge}
}

\newglossaryentry{property}{
  name=property,
  description={A local descriptor of an entity within a connection}
}

\newcommand{\eg}{e.g.\@}
\newcommand{\ie}{i.e.\@}

\usetikzlibrary{
  shapes.geometric, arrows, bending, calc, spath3, positioning
  }

\lstdefinelanguage{Lisp}{
  morekeywords={defun, lambda, let, if, cond, car, cdr, cons, list},
  sensitive=true,
  morecomment=[l]{;},
  morestring=[b]"
}

\acrodef{aar}[AAR]{Address-addressable read}
\acrodef{ai}[AI]{artificial intelligence}
\acrodef{asoca}[ASOCA]{Associative Chip Architecture}
\acrodef{asoca1}[ASOCA1]{Associative Memory Chip I}
\acrodef{asoca2}[ASOCA2]{Associative Memory Chip II}
\acrodef{asoca3}[ASOCA3]{Associative Memory Chip III}
\acrodef{car}[CAR]{Content-addressable read}
\acrodef{cc}[CC]{Central controller}
\acrodef{csv}[CSV]{comma-separated values}
\acrodef{ctpt}[CTPT]{counterpart}
\acrodef{dag}[DAG]{directed acyclic graph}
\acrodef{dam}[DAM]{dually-addressable memory}
\acrodef{drlg}[DRLG]{directed recursively labeled graph}
\acrodef{drlh}[DRLH]{directed recursive labelnode hypergraph}
\acrodef{dsm}[DSM]{distributed shared memory}
\acrodef{eoc}[\texttt{EOC}]{end of chain}
\acrodef{fol}[FOL]{first-order logic}
\acrodef{fpga}[FPGA]{field-programmable gate array}
\acrodef{gdb}[GDB]{graph database}
\acrodef{id}[ID]{identifier number}
\acrodef{rlg}[RLG]{recursively labelable graph}
\acrodef{isa}[ISA]{instruction set architecture}
\acrodef{kb}[KB]{Knowledge base}
\acrodef{ldbc}[LDBC]{Linked Data Benchmark Council}
\acrodef{lpg}[LPG]{labeled property graph}
\acrodef{nlp}[NLP]{natural language processing}
\acrodef{rag}[RAG]{retrieval-augmented generation}
\acrodef{rdf}[RDF]{Resource Description Framework}
\acrodef{rram}[ReRAM]{resistive random-access memory}
\acrodef{sf}[SF]{scale factor}
\acrodef{snb}[SNB]{Social Network Benchmark}
\acrodef{vsa}[VSA]{Vector symbolic architectures}

\tikzstyle{vertex} = [
  rectangle, minimum width=1.5cm, minimum height=0.5cm, text centered, draw=black
]
\tikzstyle{ndot} = [
  circle, yshift=-0.2cm, minimum width=0.1cm, minimum height=0.1cm, draw=black, fill=orange!30
]
\tikzstyle{data} = [
  vertex, rounded corners
]
\tikzstyle{pointer} = [
  data
]
\tikzstyle{primid1} = [
  pointer, xshift=1cm
]
\tikzstyle{primid2} = [
  pointer, xshift=-1cm
]
\tikzstyle{head} = [
  data, fill=red!30
]
\tikzstyle{prop} = [
  data, fill=green!30
]
\tikzstyle{next} = [
  data, fill=cyan!30
]
\tikzstyle{address} = [
  thick, ellipse, draw=black, fill=orange!30
]
\tikzstyle{pntrlink} = [
  thick, ->, >=stealth
]
\tikzstyle{proplink} = [
  thin, ->, >=stealth
]
\tikzstyle{nextptr} = [
  pntrlink, draw=cyan!70
]
\tikzstyle{propdot1} = [
  circle, xshift=0.3cm, yshift=-1cm, minimum width=0.1cm, minimum height=0.1cm, draw=black, fill=green!30
]
\tikzstyle{propdot2} = [
  circle, xshift=-0.3cm, yshift=-1cm, minimum width=0.1cm, minimum height=0.1cm, draw=black, fill=green!30
]

\begin{document}
\title{\textit{Views}: A Hardware-Aware Recursively Labeled Graph Database Model for Knowledge Representation and Reasoning}
\author{Yanjun Yang$^{1,*}$, Adrian Wheeldon$^{2}$, Yihan Pan$^{1}$ and Alex Serb$^{1}$}
\date{}
\maketitle
\begin{center}
\small
$^{1}$Centre for Electronics Frontiers, Institute for Integrated Micro and Nano Systems,\\
School of Engineering, University of Edinburgh, Edinburgh, United Kingdom\\
$^{2}$Literal Labs, Newcastle upon Tyne, United Kingdom\\
$^{*}$Correspondence: \href{mailto:yanjun.yang@ed.ac.uk}{yanjun.yang@ed.ac.uk}
\end{center}
\begin{center}
\small
Author's accepted manuscript. Accepted by \emph{Frontiers in Artificial Intelligence}\\
on 8 September 2026. Journal article: \url{https://doi.org/10.3389/frai.2026.1874897}.\\
This arXiv copy precedes the final publisher-formatted version.\\
\copyright\ 2026 Yang, Wheeldon, Pan and Serb.
Licensed under \href{https://creativecommons.org/licenses/by/4.0/}{CC BY 4.0}.
\end{center}

\begin{abstract}
Knowledge representation remains a central challenge for reasoning-centered artificial intelligence, particularly when semantic structures involve relations over relations, contextual annotations, and recursively nested descriptions.
This paper introduces \textit{Views}, a recursively labeled \ac{gdb} model designed to represent such graph-structured knowledge within a uniform graph abstraction while retaining a hardware-aware organization for associative search and traversal.
The model refactors directed labeled graphs into linked-list-like chains of linknodes, supports recursive labeling of vertices and edges, and admits mappings from \ac{rdf}- and \ac{lpg}-style graph representations.
We describe the data structure, its hardware-oriented memory mappings under the \ac{asoca}, and selected associative operations for retrieval over \textit{Views}-based \acp{gdb}.
We then evaluate storage footprint under stated mapping and store boundaries using three scales of the Social Network Benchmark published by the Linked Data Benchmark Council, and characterize directed K-hop traversal on an \ac{asoca3} \ac{fpga} implementation.
The storage results show that allocation and entry width materially affect the reported footprints rather than establishing an intrinsic advantage for \textit{Views}, while the K-hop measurements show increasing mean returned-vertex count and mean latency with hop bound over the fixed resident graph image.
Worked semantic-reasoning and Copycat-inspired examples further illustrate how retrieval operations can be composed over the model, rather than providing application-level reasoning or cognitive validation.
Taken together, these results position \textit{Views} as a model--architecture co-design for associative storage and selected graph traversal under the stated conditions; broader database workloads and end-to-end reasoning applications remain to be evaluated.

\par\smallskip\noindent\textbf{Keywords:} graph database, knowledge representation, recursive labeling, associative memory, graph traversal, near-memory computing
\end{abstract}

\section{Introduction}
\label{sec:intro}
\Acfp{gdb} are used to describe, organize and manipulate data in the form of graphs, while also using graph-based integrity checks. 
This endorses efficient storage, retrieval and visualization of inter-related data, especially when the workload requires frequent ``traversals'' of the graph, \ie, the sequential retrieval of related, or ``connected'', data. 
\acp{gdb} are frequently seen in areas where data connections have the same or higher importance than the data entries themselves, from industrial data analysis, such as supplier-client relationships in a supply chain, to symbolic \ac{ai}, semantic computing, and \ac{rag}, where knowledge of data inter-relationships takes a critical role in implementation.

Widespread usage of \acp{gdb} is seen in knowledge representation, particularly in the implementation of knowledge graphs. 
These are most commonly represented as \ac{rdf} or \acp{lpg} and stored in their \ac{gdb} systems\citep{hogan2021knowledge}.
\ac{rdf} represents knowledge as ``subject-predicate-object'' triples, with optional reification and named graphs for a clean logical footing \citep{d2010semantic,candan2001resource}. 
\acp{lpg}, on the other hand, attach labels and properties directly to vertices and edges for application-centric schemas and analytics. 
Across both families, the explicitness of knowledge graph semantics shows promise in symbolic cognitive computing by exposing symbols and relations to reasoning procedures \citep{langley2009cognitive,opdahl2022semantic}. 
Knowledge graphs are also being investigated as retrieval substrates for large-language-model pipelines \citep{gao2023retrieval}.
The present work concerns the underlying \ac{gdb} model and traversal substrate; it does not evaluate retrieval quality, generation quality or an end-to-end \ac{rag} pipeline.
These developments call for efficient graph data storage and processing methods to handle the growing size, heterogeneity and complexity of semantic information.

Operationally, querying is the critical workload in \acp{gdb}. 
For example, a user may ask ``find me all films which Tom Hanks has acted in'', followed by ``who is Tom Hanks'', as one might do in a film database. 
Multiple query languages and formal frameworks support expressive graph pattern matching and path traversal, including SPARQL for \ac{rdf} and GQL or Cypher for property graphs.
However, these high-level languages are built upon computing facilities which are not necessarily optimized for graph-native or reasoning-native execution \citep{angles2017foundations,barcelo2013querying}. 
Contemporary solutions such as GPU clusters can alleviate some \ac{gdb} workload throughput limits, although the cost of power, data movement and deployment complexity constrains their widespread adoption \citep{dominguez2010survey}.

This is not to say that existing \ac{gdb} models are poorly engineered. 
On the contrary, \ac{rdf} stores, \ac{lpg} systems and their associated query engines have benefited from decades of development under conventional computing assumptions. 
When executed on conventional von Neumann architectures, a newly proposed \ac{gdb} model should not be expected to immediately and universally outperform mature software systems that have been highly optimized for such platforms. 
In symbol-heavy applications, such as knowledge graphs and semantic reasoning, performance bottlenecks often arise from irregular access patterns, pointer chasing, poor cache behavior and memory bandwidth pressure rather than from arithmetic complexity alone. 
Heterogeneity of data types and cross-database operability further present difficulties in efficient processing and scaling \citep{patil2018survey,wylot2018rdf}, for example in biological science \citep{yoon2017use}.
Memory and schema optimizations are regarded as promising ways to enhance storage efficiency and query performance \citep{yoon2017use,sun2015sqlgraph,martinez2012exchange}, yet the joint design of \ac{gdb} models and graph-processing hardware remains comparatively underexplored.

The perspective of this paper is therefore \textbf{not} that a new \ac{gdb} model should simply replace established \ac{rdf} or \ac{lpg} systems in their existing software environments. 
Rather, we examine one \ac{gdb} model designed jointly with dedicated hardware for storing, searching and traversing symbolic and semantic information.
This motivates a model-architecture co-design approach, where the structure of the data model is selected not only for expressiveness, but also for its natural mapping to memory arrays, associative lookup and near-memory graph operations.
Within \textit{Views}, the graph representation defines the stored structures, while the associative execution model defines retrieval operations over them.
The \ac{asoca3} \ac{fpga} is the measured hardware implementation.
Reasoning procedures form a separate layer above the representation and operations.

Historically, recursive labeling in \ac{gdb} models has been proposed as a method to represent complex relationships in graphs, especially in semantic contexts. 
Pratt developed an initial hierarchical \ac{gdb} model of recursive labeling for the semantics of programming languages \citep{pratt1969hierarchical}, and Boley proposed \acf{drlh} as a representation language for semantic network representations, capable of being specialized into natural language or other languages such as Lisp \citep{boley1977directed}.
These methods are highly relevant to the representation of semantic information, because many relationships are not merely binary edges between entities. 
Edges can have properties, properties can themselves require further properties, and contextual information often needs to be attached locally rather than globally. 
However, these earlier models were not developed into modern database-level operational methods, nor were they directly connected to hardware implementation.

Against this background, we investigate specialized hardware that stores a graph-oriented representation and performs in- or near-memory operations over it.
This co-design permits storage footprint and traversal kernels to be examined under explicit implementation assumptions.

In this work, we propose \textit{Views}, a hardware-aware \ac{gdb} model for storing semantic information.
\textit{Views} is based on \acf{drlg}, where vertices and edges may both be labeled, and where labels themselves may have further labels to any finite depth permitted by the stored graph.
The resulting model is designed to represent entities, relations, properties, contextual annotations and higher-order semantic structures within a uniform graph abstraction. 
At its foundation, \textit{Views} refactors descriptions of directed and labeled graphs as linked-list-like structures for better uniformity and symmetry in the data structure. 
This is not merely a software-level reformulation: the structure of \textit{Views} is designed to be naturally instantiated in hardware, particularly within the \ac{asoca} program.

ASOCA provides the target architectural context for the model. 
It aims to turn the ideas behind \textit{Views} into a series of hardware accelerators for \ac{gdb} operations, using dually-addressable and content-addressable memory structures to support graph storage, associative search and near-memory traversal. 
Under this premise, the value of \textit{Views} should be understood as a model-architecture co-design. 
Its novelty lies not in claiming general superiority over mature \ac{gdb} models on conventional architectures, but in aligning recursively labeled semantic graph structures with the memory organization and associative operations targeted by ASOCA-style hardware.

This paper makes the following contributions. 
First, it introduces \textit{Views}, a recursively labeled \ac{gdb} model for encoding entities, relations, annotations and higher-order semantic structures within a uniform graph abstraction.
Second, it shows how \textit{Views} can represent conventional graph structures while supporting recursive labelability for semantic and contextual information. 
Third, it defines selected associative-memory operations used by the hardware mapping and worked retrieval examples.
Fourth, it presents \textit{Views} as part of a model-architecture co-design with ASOCA, an actively developed hardware architecture for associative graph processing. 
Fifth, it characterizes storage footprint at three Social Network Benchmark scales and directed, up-to-$k$ traversal on the later \ac{asoca3} \ac{fpga} implementation.
Finally, it uses semantic-inference and Copycat-inspired examples as representational case studies showing how retrieval primitives can be composed.

The remainder of this paper is organized as follows. 
Section 2 describes the proposed \textit{Views} \ac{gdb} model in terms of data encoding, recursive labeling and mappings from conventional graph representations.
Section 3 presents the hardware-aware implementation methodology, including mapping schemes and the ASOCA implementation. 
Section 4 reports the scaled storage study, the \ac{asoca3} K-hop traversal evaluation, an illustrative symbolic retrieval and a Copycat-inspired mapping \citep{hofstadter1995fluid}.
Section 5 discusses the relationship of \textit{Views} to other graph models, its limitations and potential extensions.

\section{Materials and Methods: The \textit{Views} GDB Model}
\label{sec:arch}

The proposed \textit{Views} \ac{gdb} model has been designed to represent graphs with the following features:
(1) \textit{Directedness}: vertex pairs are connected by directed edges (or ``arcs''),
(2) \textit{Labelability}: edges and vertices can both be ``labeled'', \ie, we can attach further properties to them and
(3) \textit{Recursive properties}: properties themselves can have properties to any finite depth represented in the stored graph, allowing complex, nested relationships.
We call these \acp{drlg}, a term inspired by Boley's \ac{drlh} \citep{boley1977directed} but restricting the underlying structure to graphs rather than hypergraphs, and any mention of ``graph'' in the rest of this paper will refer to a \ac{drlg} unless otherwise stated.

We shall now introduce the basic data structure of the proposed \ac{gdb} model and then explain its support for finite recursive labelability and its mapping to hardware.
The construction begins by identifying independently addressable concepts and encoding each source-owned relation as a linknode.
Context-dependent descriptions are attached as subordinate chains, and information is retrieved by chain traversal or associative lookup.
The section concludes by applying this flow to a compact worked example and then expanding it into a small \textit{Views}-based \ac{gdb}.

\subsection{The \textit{Views} Triplet: Mapping a ``half-\texorpdfstring{$K_2$}{K2}'' Graph}
\label{sec:unification}

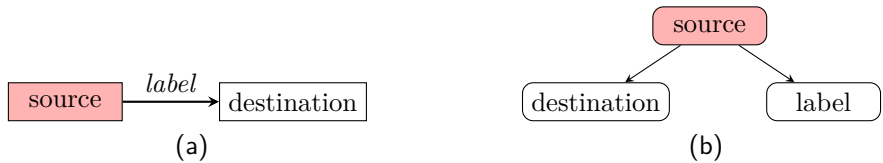
\begin{figure}[!htb]
\centering
\subfloat[]{
  \begin{tikzpicture}
    \node (source) [vertex, fill=red!30] {source};
    \node (destination) [vertex, right of=source, xshift=2cm] {destination};

    \draw [pntrlink] (source) -- node[anchor=south] {\textit{label}} (destination);
  \end{tikzpicture}
  \label{fig:ternary}
}
\hfil
\subfloat[]{
  \begin{tikzpicture}
    \node (source) [head] {source};
    \node (destination) [data, below of=source, xshift=-1.5cm] {destination};
    \node (lbl) [data, below of=source, xshift=1.5cm] {label};

    \draw [proplink] (source) -- (destination);
    \draw [proplink] (source) -- (lbl);
  \end{tikzpicture}
  \label{fig:triplet}
}
  \caption{
    Forms of basic data structure: (a) A ``half-$K_2$'' graph. Rectangles represent abstract graph vertices and the arrow represents an abstract graph edge. (b) The triplet in \textit{Views} \ac{gdb} model. Here, the bevelled rectangles represent data that can be stored in physical memory entries as numbers or pointers. 
  }
\end{figure}

We begin construction of the \textit{Views} data structure by considering a simple, non-trivial \ac{drlg} as illustrated in \Cref{fig:ternary}, which consists of two vertices (denoted in non-beveled rectangles) and a single, labeled, directed edge connecting them. 
We call this a ``half-$K_2$'' digraph (to distinguish it from the full $K_2$ digraph which includes the reciprocal edge) and treat it as the ``unit of relational information'' within a graph: repeated instantiation of such ternary relationships can be used to create any \ac{drlg}. 
In the semantic web context, this corresponds to a \ac{rdf} triple, and repeated instantiation constructs an \ac{rdf} graph \citep{d2010semantic,candan2001resource}. 
Within the form of \ac{rdf} triple, \textit{<subject-predicate-object>}, \textit{subject} corresponds to the \textit{``source'' vertex}, \textit{predicate} to the \textit{labeled edge}, and \textit{object} to the \textit{``destination'' vertex}.
Finally, note that the terms ``source'', ``edge'' and ``destination'' are used in a ``de re'' fashion and refer to the objects in question directly: we have not yet made use of labelability so far.

\textit{Views} represents the conjugate ternary relationship using a ``source-centered'', numerical structure that we shall call the \textit{Views} \textit{triplet}, meaning that \textit{edges} and \textit{destination vertices} are treated as equivalent entities (\ie, we make no distinction between them), both of which relate to the source vertex in the same fashion (as indicated by the ``-t' suffix's shape).
This is illustrated in \Cref{fig:triplet}, where the source vertex ``points to'' both the edge and the destination vertex.
Note that bevelled rectangles in the discourse differ from conventional vertices and they refer to physical memory entries as numbers or pointers, and that the arrows in this figure denote the concept of ``owns'', \ie, the source vertex ``owns'' the marked destination vertex and edge.
Thus, the pair of vertices and connecting edge have been mapped onto a set of 3 numbers that can be stored in a physical memory.
So long as there is a way to track the association between the members of the \textit{Views} triplet, we can represent a triplet in physical memory (more details in subsequent sections).

\subsection{From Triplet to ``Simple'' Vertex-labeled GDB}
\label{sec:basic_GDB}

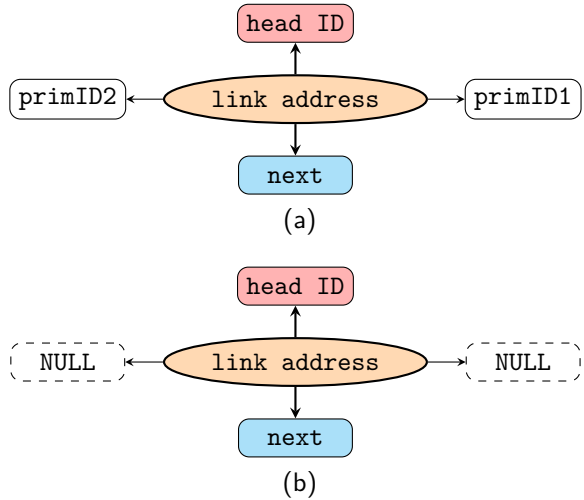
\begin{figure}[!htb]
\centering
\subfloat[]{
  \begin{tikzpicture}[scale=0.9, transform shape]
    \node (triplet) [address] {\texttt{link address}};
    \node (head) [head, above of=triplet] {\texttt{head ID}};
    \node (pid1) [pointer, right of=triplet, xshift=2.5cm] {\texttt{primID1}};
    \node (pid2) [pointer, left of=triplet, xshift=-2.5cm] {\texttt{primID2}};
    \node (next) [next, below of=triplet] {\texttt{next}};
    
    \draw [pntrlink] (triplet) -- (head);
    \draw [proplink] (triplet) -- (pid1);
    \draw [proplink] (triplet) -- (pid2);
    \draw [pntrlink] (triplet) -- (next);
  \end{tikzpicture}
  \label{fig:proto_linknode}
}
\hfil
\subfloat[]{
  \begin{tikzpicture}[scale=0.9, transform shape]
    \node (triplet) [address] {\texttt{link address}};
    \node (head) [head, above of=triplet] {\texttt{head ID}};
    \node (pid1) [pointer, dashed, right of=triplet, xshift=2.5cm] {\texttt{NULL}};
    \node (pid2) [pointer, dashed, left of=triplet, xshift=-2.5cm] {\texttt{NULL}};
    \node (next) [next, below of=triplet] {\texttt{next}};
    
    \draw [pntrlink] (triplet) -- (head);
    \draw [proplink] (triplet) -- (pid1);
    \draw [proplink] (triplet) -- (pid2);
    \draw [pntrlink] (triplet) -- (next);
  \end{tikzpicture}
  \label{fig:proto_headnode}
}
\caption{
  \textit{Views} \ac{gdb} model (a) proto-linknode, and (b) proto-headnode. 
  Ellipses in this figure refer to the physical addresses corresponding to each portrayed link/headnode, but are not explicitly stored in each linknode's allocated memory space.
  Bevelled rectangles are data explicitly stored in the linknode's allocated memory. 
  Therefore, \texttt{link address} diagrammatically represents the address of the current linknode and \texttt{next} is a physically stored pointer to the next linknode addresses while \texttt{head ID} in red is another physically stored pointer to the source vertex. 
  Note that in the case of a headnode, \texttt{head ID} stores the same value as \texttt{link address} i.e. it points to itself.
  This also relates back to \Cref{fig:triplet}. 
}
\label{fig:proto_nodes}
\end{figure}

It is easy to see that multiple instances of the numerical triplet structure above can collectively form a \ac{gdb}.
To begin building a \ac{gdb} with this structure we begin by adding multiple edges to some source vertex and to do this in practice we upgrade the \textit{Views} triplet from above into a simplified form of the data structure that \textit{Views} relies upon (the ``linknode''): the \textbf{proto-linknode}.
This ``proto-linknode'' stores the unlabeled triplet information from the previous section, mapping it to a set of three \acp{id}. 
Internally, a proto-linknode is formatted as follows: [``source vertex'', ``edge'', ``destination vertex'', ``next linknode''], abbreviated to: [\texttt{head ID}, \texttt{primID1}, \texttt{primID2}, \texttt{next}], where primID stands for ``primary ID'', illustrated in \Cref{fig:proto_linknode}. 
\texttt{head ID} corresponds to the ``source vertex'' whilst \texttt{primID1} and \texttt{primID2} correspond to the ``edge'' and ``destination vertex'' respectively.
Note how the updated naming convention makes it explicitly clear that there is no distinction in principle between edge and destination vertex. 

With the linknode partly defined we can now add multiple linknodes with the same \texttt{head ID} in order to add further properties to some source vertex; each corresponds to an outgoing edge with its label and destination vertex of the source vertex in a graph.
The collection of all linknodes ``belonging to'' some object X is called the \textbf{chain} of X.
What all linknodes in a chain have in common is that they share \texttt{head ID}.
The \texttt{next} pointer (at the bottom of \Cref{fig:proto_linknode}) is added to allow the \ac{gdb} to ``traverse'' (sequentially traverse) a chain, even if its linknodes are not  stored sequentially in memory (\ie, the chain is ``fragmented'').
Starting from the head of a chain, \texttt{next} points to another linknode belonging to the same source vertex, and effectively turns a chain into a traversable linked list.
Thus, an object with $N$ properties will feature $N$ linknodes arranged into a linked list.
The objective here is to allow a near-memory processor to autonomously ``discover'' all linknodes attached to a particular source vertex/\texttt{head ID} without having to query the entire memory (\eg, via broadcast), which could be extremely large.
Compare the cost of energising 32 billion memory entries to check if they are of interest, to following a couple of hundred linknodes by ``hopping'' from one to the other using the \texttt{next} pointer.
Thus, the implementation of a \texttt{next} pointer is very much rooted in hardware considerations.
We have now explained the entire structure of the proto-linknode.

The question now arises: when forming a new chain how should the linknodes be ordered in the linked list and particularly, who should be first in the list?
In this version of \textit{Views}, we define a special linknode flavour called \textbf{headnode} (its primitive ``proto'' version is shown in \Cref{fig:proto_headnode}) and we do not impose any particular requirements on the ordering of the rest of the linknodes.
Intuitively we can think of a headnode as stating that ``the object at \texttt{[link address]} exists as an entity''.
Practically, it acts as the origin of X's chain.
The headnode has the exact same structure as any other linknode; the difference is in its contents.
Within headnodes exclusively: \texttt{head ID} points to its own address (link address) and both primIDs are empty (\texttt{NULL}).
The headnode's self-reference distinguishes it from linknodes as a discrete source vertex. 
Finally, we note that to end a chain, we fill the final linknode's \texttt{next} pointer to a specially chosen value that represents the \ac{eoc}.
\ac{eoc} defines the end point of a chain to effectively terminate traversal operations. 
This is similar to the usage of end-of-file markers in file systems.
Therefore, all chains in \textit{Views} are finite in runtime by the \ac{eoc} and can be traversed in a linear fashion.

\begin{figure*}[!htb]
\centering
\subfloat[]{
  \begin{tikzpicture}
    \node (0x00a) [vertex, fill=red!30] {\texttt{0x00a}};
    \node (black) [vertex, below of=0x00a, yshift=-0.25cm, xshift=-2.5cm] {black};
    \node (naughty) [vertex, below of=0x00a, yshift=-0.25cm, xshift=2.5cm] {naughty};
    \node (cat) [vertex, above of=0x00a, yshift=0.5cm] {cat};

    \draw [pntrlink] (0x00a) -- node[anchor=west] {\textit{color}} (black);
    \draw [pntrlink] (0x00a) -- node[anchor=west] {\textit{temper}} (naughty);
    \draw [pntrlink] (0x00a) -- node[anchor=west] {\textit{species}} (cat);
  \end{tikzpicture}
  \label{fig:graph_sentence}
}
\hfil
\subfloat[]{
  \begin{tikzpicture}[scale=0.9, transform shape]
    \node (0x00a) [data, ellipse, fill=red!30] {\texttt{0x00a}};

    \node (null1) [primid1, dashed, xshift=0.5cm, right of=0x00a] {NULL};
    \draw [proplink] (0x00a) -- (null1);
    \node (null2) [primid2, dashed, xshift=-0.5cm, left of=0x00a] {NULL};
    \draw [proplink] (0x00a) -- (null2);

    \node (TempNaugh) [ndot, ellipse, below of=0x00a] {\texttt{0x01d}};
    \draw [nextptr] (0x00a) -- (TempNaugh);
    \node (temper) [primid1, xshift=0.5cm, right of=TempNaugh] {temper};
    \draw [proplink] (TempNaugh) -- (temper);
    \node (naughty) [primid2, xshift=-0.5cm, left of=TempNaugh] {naughty};
    \draw [proplink] (TempNaugh) -- (naughty);

    \node (ColBla) [ndot, ellipse, below of=TempNaugh] {\texttt{0x00b}};
    \draw [nextptr] (TempNaugh) -- (ColBla);
    \node (color) [primid1, xshift=0.5cm, right of=ColBla] {color};
    \draw [proplink] (ColBla) -- (color);
    \node (black) [primid2, xshift=-0.5cm, left of=ColBla] {black};
    \draw [proplink] (ColBla) -- (black);

    \node (SpecCat) [ndot, ellipse, below of=ColBla] {\texttt{0xfed}};
    \draw [nextptr] (ColBla) -- (SpecCat);
    \node (species) [primid1, xshift=0.5cm, right of=SpecCat] {species};
    \draw [proplink] (SpecCat) -- (species);
    \node (cat) [primid2, xshift=-0.5cm, left of=SpecCat] {cat};
    \draw [proplink] (SpecCat) -- (cat);

    \node (eoc) [next, below of=SpecCat, fill=orange!30] {\ac{eoc}};
    \draw [nextptr] (SpecCat) -- (eoc);
  \end{tikzpicture}
  \label{fig:chain_sentence}
}
\caption{
  A semantic sentence ``Object \texttt{0x00a} is a naughty black cat'' equivalently stored in: (a) a graph where a vertex has degree 3, (b) a \textit{Views} chain with length 4.
  Note that the headnode link address oval has been colored red to highlight that it is a headnode.
  \ac{eoc} is a special value used to indicate the end of a chain instead of valid linknodes. 
}
\label{fig:graph2chain}
\end{figure*}
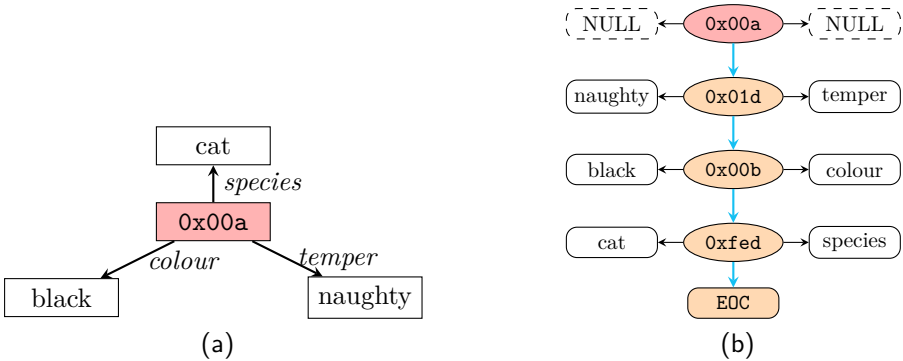

With the ability to create chains, we can now continue building our \ac{gdb}.
We begin by noting that -- crucially -- in this version of \textit{Views} primIDs from any linknode point to headnodes.
This is how connections are made from a chain to other chains and how a source vertex is connected to a destination vertex via an edge in practice.
Other versions of \textit{Views} may relax that requirement and impose ordering criteria on the formation of the linked list, but these lie beyond the scope of this paper.
Examining \Cref{fig:proto_nodes} we note that we now know exactly how to populate all physical fields of any number of headnodes and linknodes that we may have in the database.
To illustrate this, \Cref{fig:graph2chain} shows how a source vertex with 3x (directed) connections maps to a 4-node chain.

We note that due to the presence of the headnode a vertex of degree $\delta$ maps to a chain of length $\delta$+1:
\begin{equation}
  l(v) = \delta (v) + 1
  \label{eq:degree2length}
\end{equation}
where $\delta (v)$ is the degree of vertex $v$, and $l(v)$ is the length of the linked list of vertex $v$.

\subsection{The full \textit{Views} Linknode: Adding Labelability}
\label{sec:linknode}

\begin{figure}[!htb]
\centering
\subfloat[]{
  \begin{tikzpicture}
    \node (triplet) [address] {\texttt{link address}};
    \node (head) [head, above of=triplet, yshift=0.75cm] {\texttt{head ID}};
    \node (pid1) [pointer, right of=triplet, xshift=1cm, yshift=0.75cm] {\texttt{primID1}};
    \node (pid2) [pointer, left of=triplet, xshift=-1cm, yshift=0.75cm] {\texttt{primID2}};
    \node (prop1) [prop, right of=triplet, xshift=1cm, yshift=-1.5cm, text width=1.5cm] {\texttt{prop1}};
    \node (prop2) [prop, left of=triplet, xshift=-1cm, yshift=-1.5cm, text width=1.5cm] {\texttt{prop2}};
    \node (next) [next, below of=triplet, yshift=-0.75cm] {\texttt{next}};
    
    \draw [pntrlink] (triplet) -- (head);
    \draw [proplink] (triplet) to [out=60, in=180] (pid1);
    \draw [proplink] (triplet) to [out=120, in=0] (pid2);
    \draw [proplink] (triplet) to [bend left=30] (prop1);
    \draw [proplink] (triplet) to [bend right=30] (prop2);
    \draw [pntrlink] (triplet) -- (next);
  \end{tikzpicture}
  \label{fig:linknode}
}
\hfil
\subfloat[]{
  \begin{tikzpicture}
    \node (triplet) [address] {\texttt{link address}};
    \node (head) [head, above of=triplet, yshift=0.75cm] {\texttt{head ID}};
    \node (pid1) [pointer, dashed, right of=triplet, xshift=1cm, yshift=0.75cm, text width=1.5cm] {\texttt{NULL}};
    \node (pid2) [pointer, dashed, left of=triplet, xshift=-1cm, yshift=0.75cm, text width=1.5cm] {\texttt{NULL}};
    \node (prop1) [prop, right of=triplet, xshift=1cm, yshift=-1.5cm, text width=1.5cm] {\texttt{head\\prop}};
    \node (prop2) [prop, dashed, left of=triplet, xshift=-1cm, yshift=-1.5cm, text width=1.5cm] {\texttt{NULL}};
    \node (link) [next, below of=triplet, yshift=-0.75cm] {\texttt{next}};

    \draw [pntrlink] (triplet) -- (head);
    \draw [pntrlink, dashed] (head) to [bend right=45] (triplet);
    \draw [proplink] (triplet) to [out=60, in=180] (pid1);
    \draw [proplink] (triplet) to [out=120, in=0] (pid2);
    \draw [proplink] (triplet) to [bend left=30] (prop1);
    \draw [proplink] (triplet) to [bend right=30] (prop2);
    \draw [pntrlink] (triplet) -- (link);
  \end{tikzpicture}
  \label{fig:headnode}
}
\caption{
  \textit{Views} \ac{gdb} model (a) linknode, and (b) headnode taking the same format from \Cref{fig:proto_nodes}. 
  \texttt{prop1} and \texttt{prop2} are supplemented into our data structure. 
  Note that the properties of the source vertex itself are stored in the location of \texttt{prop1} in its headnode, to which its head ID points. 
}
\end{figure}
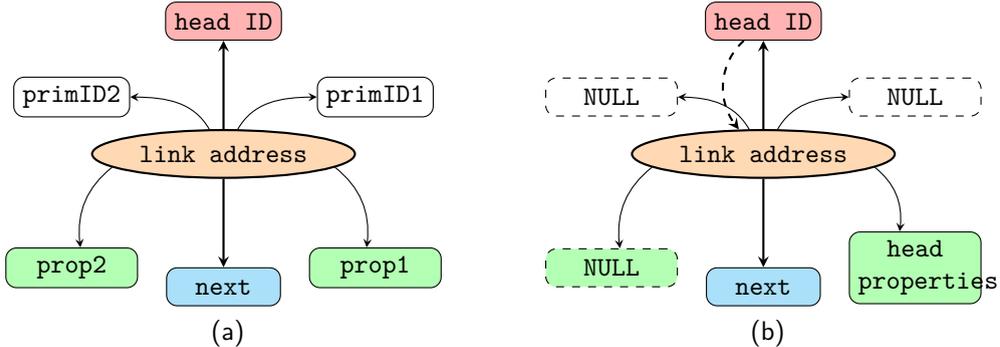

We now extend our proto-linknode structure to support recursive properties and become \textbf{traversable} in an elegant way.
This upgrades the basic structure described in \Cref{fig:ternary} into what we see in \Cref{fig:linknode}: the full \textit{Views} \textbf{linknode} data structure.
Internally, a linknode is formatted as follows: [``source vertex'', ``edge'', ``edge properties'', ``destination vertex'', ``destination vertex properties'', ``next linknode''], abbreviated to: [\texttt{head ID}, \texttt{primID1}, \texttt{prop1}, \texttt{primID2}, \texttt{prop2}, \texttt{next}], where primID stands for ``primary ID'' and ``prop'' for ``properties'', illustrated in \Cref{fig:linknode}.
The corresponding structure of headnode format is illustrated in \Cref{fig:headnode}.

In the previous section, we have covered the ``labeling'' process for source vertices, but edges and destination vertices have not yet been ``labeled''.
To do so, we furnish the \textit{edge} and \textit{destination vertex} entities with \textbf{relationship-specific} properties, illustrated in \Cref{fig:linknode} at the lower left and lower right (\texttt{prop1} and \texttt{prop2}), to enable recursive labelability.
As before, these are also numbers stored in a physical memory and each number corresponds to an \textit{address in memory}, \ie, they are \textbf{pointers}.
They point to linknodes that elaborate on the properties of \texttt{primID1} and \texttt{primID2} \textit{within the context of the linknode's triplet, \ie, within the context of the relationship as a whole}.
The ``context-free'' versus ``context-dependent'' distinction is critical:
In abstract terms it denotes the difference between ``a property of object X'' and ``a property of object X \textit{when it relates to Y via Z}''.

In the schema this model was conceived for, a primID is essentially an identifier number as a pointer, but it can point to either an independent object (\eg, ``that chestnut brown bird on this green tree'') or a concept (or class, ``sparrow'' and ``camphor tree''). 
Or in short, a primID refers to a specific entity that is described by its own linknodes.
prop pointers then handle context-specific information, which has no meaning of its own outside the specific context of the relation being specified by the linknode.
This places objects and classes at the same structural level without requiring recursive inheritance to express this distinction.
It is thus possible to build a knowledge graph with uniform data structures while retaining contextual descriptions.

\tikzstyle{primid1} = [
  pointer, xshift=0.7cm, yshift=0.5cm
]
\tikzstyle{primid2} = [
  pointer, xshift=-0.7cm, yshift=0.5cm
]
\tikzstyle{pid1} = [
  pointer, minimum width=0, xshift=0.3cm, yshift=0.4cm
]
\tikzstyle{pid2} = [
  pointer, minimum width=0, xshift=-0.3cm, yshift=0.4cm
]

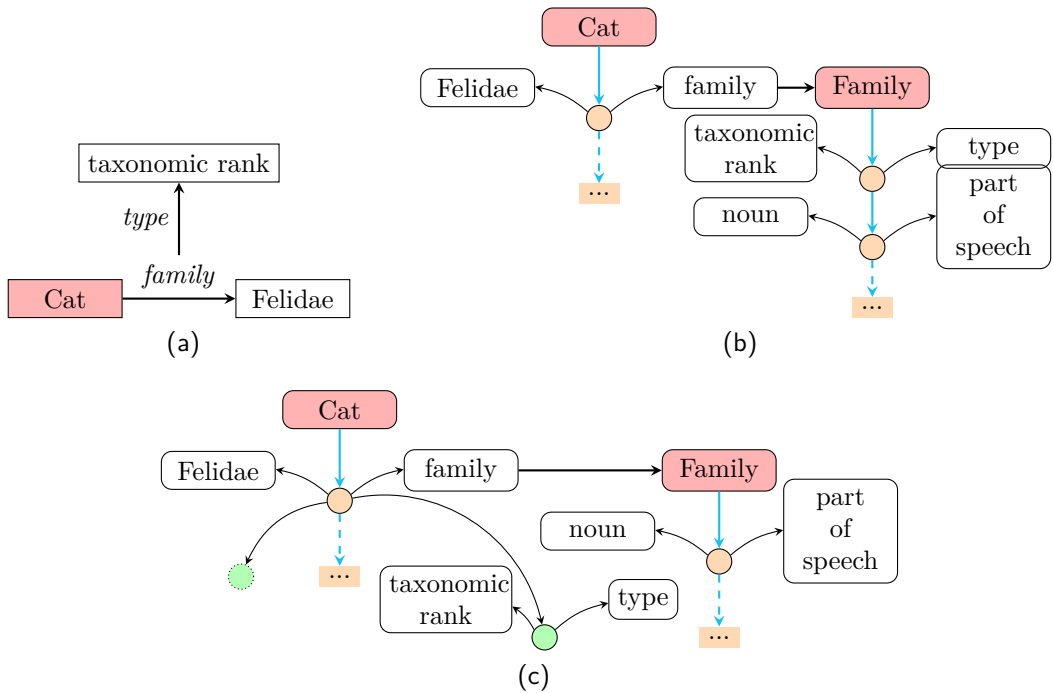
\begin{figure}[!htb]
\centering
\subfloat[]{
  \begin{tikzpicture}
    \node (cat) [vertex, fill=red!30] {Cat};
    \node (felidae) [vertex, right of=cat, xshift=2cm] {Felidae};

    \draw [pntrlink] (cat) -- (felidae) node[midway, anchor=south, name=family] {\textit{family}};

    \node (taxo) [vertex, above of=family, yshift=0.5cm] {taxonomic rank};
    \draw [pntrlink] (family) -- node[anchor=east] {\textit{type}} (taxo);

  \end{tikzpicture}
  \label{fig:nested}
}
\hfil
\subfloat[]{
  \begin{tikzpicture}
    \node (Cat) [head] {Cat};

    \node (FamilyFelidae) [ndot, below of=Cat] {};
    \draw [nextptr] (Cat) -- (FamilyFelidae);
    \node (Family) [primid1, right of=FamilyFelidae] {family};
    \draw [proplink] (FamilyFelidae) to [bend left=15] (Family);
    \node (Felidae) [primid2, left of=FamilyFelidae] {Felidae};
    \draw [proplink] (FamilyFelidae) to [bend right=15] (Felidae);

    \node (Cont1) [fill=orange!30, below of=FamilyFelidae] {...};
    \draw [nextptr, dashed] (FamilyFelidae) -- (Cont1);
    
    \node (F) [head, right of=Family, xshift=1cm] {Family};
    \draw [pntrlink] (Family) -- (F);
    \node (TypeTax) [ndot, below of=F, yshift=-0.5cm] {};
    \draw [nextptr] (F) -- (TypeTax);
    \node (Type) [primid1, right of=TypeTax] {type};
    \draw [proplink] (TypeTax) to [bend left=15] (Type);
    \node (Tax) [primid2, text width=2cm, left of=TypeTax] {taxonomic rank};
    \draw [proplink] (TypeTax) to [bend right=15] (Tax);

    \node (PosNoun) [ndot, below of=TypeTax] {};
    \draw [nextptr] (TypeTax) -- (PosNoun);
    \node (Pos) [primid1, text width=1.5cm, right of=PosNoun] {part of speech};
    \draw [proplink] (PosNoun) to [bend left=15] (Pos);
    \node (Noun) [primid2, left of=PosNoun] {noun};
    \draw [proplink] (PosNoun) to [bend right=15] (Noun);

    \node (Cont2) [fill=orange!30, yshift=0.2cm, below of=PosNoun] {...};
    \draw [nextptr, dashed] (PosNoun) -- (Cont2);
  \end{tikzpicture}
  \label{fig:taxo_family}
}
\hfil
\subfloat[]{
  \begin{tikzpicture}
    \node (Cat) [head] {Cat};

    \node (FamilyFelidae) [ndot, below of=Cat] {};
    \draw [nextptr] (Cat) -- (FamilyFelidae);
    \node (Family) [primid1, right of=FamilyFelidae] {family};
    \draw [proplink] (FamilyFelidae) to [bend left=15] (Family);
    \node (Felidae) [primid2, left of=FamilyFelidae] {Felidae};
    \draw [proplink] (FamilyFelidae) to [bend right=15] (Felidae);
    
    \node (dummy1) [propdot2, densely dotted, left of=FamilyFelidae] {};
    \draw [proplink] (FamilyFelidae) to [bend right=30] (dummy1);

    \node (Cont1) [fill=orange!30, below of=FamilyFelidae] {...};
    \draw [nextptr, dashed] (FamilyFelidae) -- (Cont1);

    \node (TypeTax) [propdot1, yshift=-1.5cm, xshift=0.5cm, right of=FamilyFelidae] {};
    \draw [proplink] (FamilyFelidae) to [bend left=45] (TypeTax);
    \node (Type) [pid1, yshift=0.2cm, right of=TypeTax] {type};
    \draw [proplink] (TypeTax) to [bend left=15] (Type);
    \node (Tax) [pid2, yshift=0.2cm, text width=2cm, left of=TypeTax, xshift=-0.3cm] {taxonomic rank};
    \draw [proplink] (TypeTax) to [bend right=15] (Tax);
  
    \node (F) [head, right of=Family, xshift=2.5cm] {Family};
    \draw [pntrlink] (Family) -- (F);
    \node (PosNoun) [ndot, below of=F, yshift=-0.5cm] {};
    \draw [nextptr] (F) -- (PosNoun);
    \node (Pos) [primid1, text width=1.5cm, right of=PosNoun] {part of speech};
    \draw [proplink] (PosNoun) to [bend left=15] (Pos);
    \node (Noun) [primid2, left of=PosNoun] {noun};
    \draw [proplink] (PosNoun) to [bend right=15] (Noun);

    \node (Cont2) [fill=orange!30, below of=PosNoun] {...};
    \draw [nextptr, dashed] (PosNoun) -- (Cont2);
  \end{tikzpicture}
  \label{fig:normal_family}
}
  \caption{
    Examples of secondary labeling in (a) a traditional directed graph, (b) a \textit{Views}-based \ac{gdb}, where the ``family'' chain has been set up to refer to the taxonomic rank specifically, and (c) another \textit{Views}-based \ac{gdb}, where the shown ``family'' chain has been set up to represent a more generic concept of the term, only specifying the part of speech.
    Note that the white ``family'' is nothing but pointer to the head ID of a linked list, within which linknodes define it further. 
  }
  \label{fig:schema}
\end{figure}

Note, however, that whilst this is a ``natural'' choice, it is not obligatory. 
In particular \textit{Views}-based \ac{gdb} instantiations we may want to reserve headnode for classes and treat all individuals as sub-chains. 
Whether that is sufficient to uniquely identify the individual in question (for example, a particular dog) is a matter for the database contents. 
This is where a database engineer is needed to determine the specific schema to be used by the database under consideration. 
The underlying data structure allows the flexibility to choose what will be treated as a headnode.

Illustrative differences can be found between \Cref{fig:taxo_family} and \Cref{fig:normal_family}, both of which describe the nested relationships in \Cref{fig:nested}: it is up to the database schema whether to treat word meanings as context-dependent properties or as separate linknodes of their own. 
The linknode on the left (shrunk to a small orange circle for brevity) tells us that ``cats belong to the family of Felidae''.
In this particular example the ``family'' primID points to the head of the chain that contains information on what a ``family'' is, independent of context.
Following the link and inspecting the chain we find that this particular ``family'' chain refers to the taxonomical context and is itself a noun.
This does not depend in any way on the fact that cats belong to the family of Felidae, and in fact can be pointed to by multiple primIDs scattered throughout the database.

In contrast, in the alternative example of \Cref{fig:normal_family} the primID ``family'' points to the head of a chain where we have chosen to represent the generic concept of a family.
In this example, this ``family'' chain informs us that it is a noun in a context-independent manner.
On the other hand, we note that we have now used \texttt{prop1} to indicate that in the context of ``cats belonging to the family of Felidae'', ``family'' refers to the taxonomical interpretation.
Beyond that point, the chain for ``family'' may have further specifications as to what the different meanings of the term imply, and the information encoded in the linknode emitted from \texttt{prop1} can be used to seek that information.

With the above in mind, we note that in recursively labeled graphs, an edge itself can be further linked to other vertices via labeled edges, and so forth \citep{angles2008survey,boley1977directed} (\Cref{fig:nested}).
In \textit{Views} this is enabled by the ``prop'' pointers and additionally destination vertices can also be labeled by in-context properties in the exact same manner as edges.

\begin{figure}[!htb]
\centering
  \subfloat[]{
    \begin{tikzpicture}[scale=0.9, transform shape]
      \node (S) [head] {This Soup};
  
      \node (ContainsChicken) [ndot, below of=S] {};
      \draw [nextptr] (S) -- (ContainsChicken);
      \node (Contains) [primid1, right of=ContainsChicken] {contains};
      \draw [proplink] (ContainsChicken) to [bend left=15] (Contains);
      \node (Chicken) [primid2, left of=ContainsChicken] {chicken};
      \draw [proplink] (ContainsChicken) to [bend right=15] (Chicken);
      
      \node (dummy1) [propdot1, densely dotted, right of=ContainsChicken] {};
      \draw [proplink] (ContainsChicken) to [bend left=30] (dummy1);
  
      \node (PartBreast) [propdot2, xshift=-1.5cm, yshift=-0.5cm, left of=ContainsChicken] {};
      \draw [proplink] (ContainsChicken) to [bend right=20] (PartBreast);
      \node (Part) [pid1, right of=PartBreast] {part};
      \draw [proplink] (PartBreast) to [bend left=15] (Part);
      \node (Breast) [pid2, left of=PartBreast] {breast};
      \draw [proplink] (PartBreast) to [bend right=15] (Breast);
  
      \node (ShapeCubes) [ndot, yshift=0.4cm, below of=PartBreast] {};
      \draw [nextptr] (PartBreast) -- (ShapeCubes);
      \node (Shape) [pid1, right of=ShapeCubes] {shape};
      \draw [proplink] (ShapeCubes) to [bend left=15] (Shape);
      \node (Cubes) [pid2, left of=ShapeCubes] {cubes};
      \draw [proplink] (ShapeCubes) to [bend right=15] (Cubes);
  
      \node (MariSoy) [ndot, yshift=0.2cm, below of=ShapeCubes] {};
      \draw [nextptr] (ShapeCubes) -- (MariSoy);
      \node (Marinated) [pid1, text width=2cm, right of=MariSoy, xshift=0.2cm] {marinated in};
      \draw [proplink] (MariSoy) to [bend left=15] (Marinated);
      \node (Soysauce) [pid2, text width=1.2cm, left of=MariSoy] {soy sauce};
      \draw [proplink] (MariSoy) to [bend right=15] (Soysauce);
  
      \node (C) [head, left of=Chicken, xshift=-3cm] {Chicken};
      \draw [pntrlink] (Chicken) -- (C);
      \node (SpeciesGg) [ndot, below of=C] {};
      \draw [nextptr] (C) -- (SpeciesGg);
      \node (Species) [primid1, right of=SpeciesGg] {species};
      \draw [proplink] (SpeciesGg) to [bend left=15] (Species);
      \node (Gg) [primid2, left of=SpeciesGg, xshift=-0.2cm, text width=1.5cm] {Gallus gallus};
      \draw [proplink] (SpeciesGg) to [bend right=15] (Gg);

      \node (eoc) [next, below of=MariSoy, fill=orange!30] {\ac{eoc}};
      \draw [nextptr] (MariSoy) -- (eoc);
      \draw [nextptr] (SpeciesGg) |- (eoc);
      \draw [nextptr] (ContainsChicken) |- (eoc);
    \end{tikzpicture}
    \label{fig:chicken_soup}
  }
  \hfil
  \subfloat[]{
    \begin{tikzpicture}[scale=0.9, transform shape]
      \node (F) [head, minimum width=0] {Film};
      \node (IsForm) [ndot, below of=F] {};
      \node (IsA3) [pid1, xshift=-0.3cm, right of=IsForm] {is a};
      \draw [proplink] (IsForm) -- (IsA3);
      \draw [nextptr] (F) -- (IsForm);
      \node (Form) [pid2, text width=1cm, left of=IsForm] {form};

      \node (dummy) [propdot1, densely dotted, xshift=-0.3cm, right of=IsForm] {};
      \draw [proplink] (IsForm) to [bend left=30] (dummy);

      \draw [proplink] (IsForm) -- (Form);
      \node (OfTelling) [propdot2, left of=IsForm, yshift=-0.5cm] {};
      \draw [proplink] (IsForm) to [bend right=30] (OfTelling);
      \node (Of) [pid1, xshift=-0.4cm, right of=OfTelling, yshift=0.1cm] {of};
      \draw [proplink] (OfTelling) -- (Of);
      \node (Telling) [pid2, text width=2cm, xshift=-1cm, left of=OfTelling] {visual storytelling};
      \draw [proplink] (OfTelling) -- (Telling);
      
      \node (ThroughSeq) [propdot2, xshift=0.3cm, left of=OfTelling, yshift=-0.5cm] {};
      \draw [proplink] (OfTelling) to [bend right=30] (ThroughSeq);
      \node (Through) [pid1, right of=ThroughSeq, yshift=0.1cm] {through};
      \draw [proplink] (ThroughSeq) -- (Through);
      \node (Seq) [pid2, text width=2cm, xshift=-1cm, left of=ThroughSeq] {a sequence};
      \draw [proplink] (ThroughSeq) -- (Seq);
      
      \node (OfImages) [propdot2, xshift=0.6cm, left of=ThroughSeq, yshift=-0.5cm] {};
      \draw [proplink] (ThroughSeq) to [bend right=30] (OfImages);
      \node (Of2) [pid1, xshift=-0.4cm, right of=OfImages] {of};
      \draw [proplink] (OfImages) -- (Of2);
      \node (Images) [pid2, text width=2cm, xshift=-1cm, left of=OfImages] {moving images};
      \draw [proplink] (OfImages) -- (Images);
      
      \node (eoc) [next, below of=IsForm, yshift=-4cm, fill=orange!30] {\ac{eoc}};
      \draw [nextptr] (IsForm) -- (eoc);
      \draw [nextptr] (OfImages) |- (eoc);
      \draw [nextptr] (ThroughSeq) -| (eoc);
      \draw [nextptr] (OfTelling) -| (eoc);
      \draw [nextptr] (dummy) |- (eoc);
      
    \end{tikzpicture}
  \label{fig:film_subs}
  }
  \caption{
    Examples of: (a) An in-context  subordinate chain of length > 1 (``This soup contains chicken, about which we know that [the part is breast, it is in cubes and it is marinated in soy sauce].'' and ``The species of name for chicken is Gallus gallus''), and
    (b) Multiple levels of in-context labeling, which reads as \textit{Film-is a-form\{of-visual storytelling[through-a sequence(of-moving images)]\}}, or simply ``Film is a form of visual storytelling through a sequence of moving images''.
    We explicitly show pointers to ``end of chain'' (EOC) to indicate that each green-labeled linknode is not part of a single chain, but rather its own subordinate chain.
  }
\end{figure}
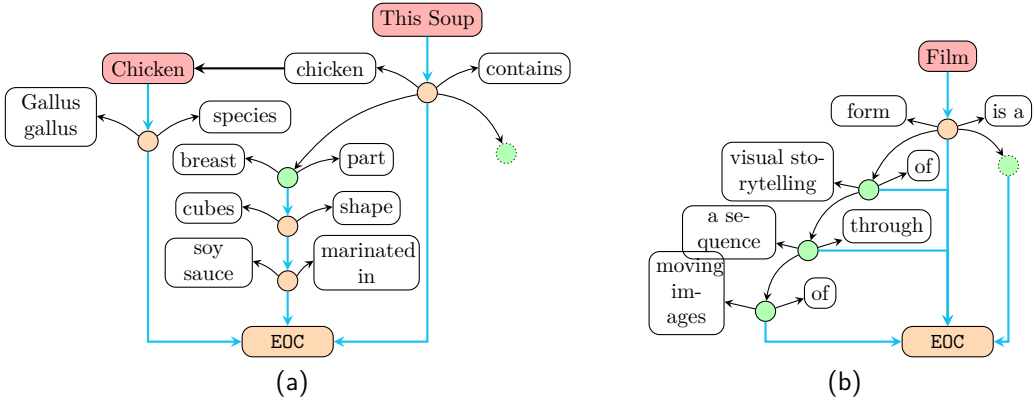

In a similar vein to chaining in \Cref{sec:basic_GDB}, successive linknodes can be emitted from \texttt{prop1} or \texttt{prop2}, forming what we shall call ``subordinate chains'' (\textbf{sub-chains}).
\Cref{fig:chicken_soup} provides an example of a subordinate chain.
The example states that a particular soup contains chicken, about which we know that the part of interest is ``breast'', and that it is chopped into cubes and marinated in soy sauce.

This information is encoded as a subordinate chain elaborating on the concept of ``chicken'' in-context.
This implements finite recursive labelability for primIDs and aids subtree search and graph mining under the proposed model \citep{jiang2013survey,tatti2024explainable}.
We make two further notes: First, a prop pointer points directly to the first linknode of a subordinate chain, and that the linknode that emits the sub-chains acts as the starting node, using the parent context for identification.
Second, a sub-chain can emit its own sub-chains recursively, as shown in \Cref{fig:film_subs}.
This emphasizes recursive labelability through finite subordinate chains, and the proposed \ac{gdb} model is therefore capable of representing complex contextual information; implementations traversing untrusted graphs require a depth limit or visited-set guard.

\subsection{Building a \textit{Views}-based Graph Database}
\label{sec:org}

Consider the conventional labeled relation
\[
  \text{Tom Hanks} \xrightarrow[\text{as: Sully}]{\text{acts in}} \text{This Film},
\]
where ``as: Sully'' annotates this particular acting relation rather than Tom Hanks or the general concept ``acts in''.
The following steps transform this relation into \textit{Views} and retrieve the annotation.

\begin{table}[!htb]
  \centering
  \caption{Constructed \textit{Views} records for the relation ``Tom Hanks acts in This Film as Sully''}
  \label{tab:worked_views}
  \small
  \begin{tabular}{@{}llll@{}}
    \toprule
    Record & Context & Primary contents & Navigation \& subordinates \\
    \midrule
    \texttt{0x1} & Tom Hanks main chain & acts in; This Film & \texttt{next}=\ac{eoc}; \texttt{prop1}$\rightarrow$\texttt{0xe} \\
    \texttt{0xe} & \texttt{0x1}, \texttt{primID1} side & as; Sully & \texttt{next}=\ac{eoc}; properties=\texttt{NULL} \\
    \bottomrule
  \end{tabular}
\end{table}

\begin{enumerate}
  \item Allocate headnodes for the independently addressable concepts \textit{Tom Hanks}, \textit{acts in}, \textit{This Film}, \textit{as} and \textit{Sully}. Each headnode begins the main chain describing its concept.
  \item Create the regular linknode at \texttt{0x1} in the \textit{Tom Hanks} main chain. Without loss of generality, \texttt{primID1} identifies the relation \textit{acts in} and \texttt{primID2} identifies the destination \textit{This Film}.
  \item Create the subordinate linknode at \texttt{0xe}, storing \textit{as} in \texttt{primID1} and \textit{Sully} in \texttt{primID2}, with \texttt{prop1} of \texttt{0x1} pointing to it. Its parent address and property side supply its context, so ``as Sully'' is not inserted into the main chain of Tom Hanks or \textit{acts in}.
  \item Apply the same construction recursively when a contextual value itself requires qualification: a property pointer of \texttt{0xe} may start another finite subordinate chain. \Cref{fig:film_subs} shows this repeated over several levels.
  \item To retrieve ``Who does Tom Hanks act as in this film?'', locate the matching \textit{acts in} linknode in the \textit{Tom Hanks} chain, follow its \texttt{prop1} pointer to \texttt{0xe}, and read the pair \textit{as}--\textit{Sully}. When the owning chain is unknown, associative lookup may locate the record first; the corresponding operations are defined in \Cref{sec:asoca}.
\end{enumerate}

\Cref{tab:worked_views} summarizes the two generated records (as a main chain and a subordinate chain in \Cref{fig:th_twin}) without imposing their physical array layout; the CNSM mapping of the same logical fields is introduced in \Cref{sec:cnsm}.
The larger example below adds further main chains and relations, while \Cref{fig:lisp_example} later compares the resulting linked organization with one possible Lisp encoding.

\tikzstyle{fakejumper} = [
  draw=white, line cap=butt, line width=0.06cm
]

\begin{figure*}[!htb]
  \centering
  \begin{tikzpicture}[scale=1.0, transform shape]
    \node (TH) [head, minimum width=0] {Tom Hanks};

    \node (ActFilm) [ndot, below of=TH] {\tiny \texttt{0x1}};
    \draw [nextptr] (TH) -- (ActFilm);
    \node (ActIn) [pid1, yshift=0.1cm, right of=ActFilm] {act in};
    \draw [proplink] (ActFilm) -- (ActIn);
    \node (ThisFilm) [pid2, left of=ActFilm, xshift=-0.2cm] {this film};
    \draw [proplink] (ActFilm) -- (ThisFilm);

    \node (AsSully) [propdot1, yshift=-0.1cm, xshift=0.9cm, right of=ActFilm] {\tiny \texttt{0xe}};
    \draw [proplink] (ActFilm) to [bend left=30] (AsSully);
    \node (As) [pid1, yshift=0.5cm, right of=AsSully] {as};
    \draw [proplink] (AsSully) -- (As);
    \node (Sully1) [pid2, xshift=0.0cm, yshift=0.0cm, left of=AsSully] {Sully};
    \draw [proplink] (AsSully) -- (Sully1);

    \node (dummy1) [propdot2, densely dotted, yshift=0.4cm, xshift=0.3cm, left of=ActFilm] {};
    \draw [proplink] (ActFilm) to [bend right=30] (dummy1);

    \node (WonOscars) [ndot, yshift=-1cm, below of=ActFilm] {\tiny \texttt{0x2}};
    \draw [nextptr] (ActFilm) -- (WonOscars);
    \node (Won) [pid1, right of=WonOscars, xshift=0.6cm, yshift=-0.3cm] {won};
    \draw [proplink] (WonOscars) -- (Won);
    \node (2Oscars) [pid2, yshift=0.15cm, left of=WonOscars] {2 Oscars};
    \draw [proplink] (WonOscars) -- (2Oscars);

    \node (dummy2) [propdot1, densely dotted, yshift=-0.3cm, xshift=-0.5cm, right of=WonOscars] {};
    \draw [proplink] (WonOscars) to [bend left=30] (dummy2);

    \node (ForBestActor) [propdot2, yshift=-2cm, left of=WonOscars] {\tiny \texttt{0xf}};
    \draw [proplink] (WonOscars) to [bend right=20] (ForBestActor);
    \node (For) [pid1, xshift=-0.6cm, yshift=0.2cm, right of=ForBestActor] {for};
    \draw [proplink] (ForBestActor) -- (For);
    \node (BestActor) [pid2, xshift=0.3cm, yshift=0.4cm, text width=1cm, left of=ForBestActor] {best actor};
    \draw [proplink] (ForBestActor) -- (BestActor);

    \node (eoc) [next, below of=TH, yshift=-7.5cm, fill=orange!30] {\ac{eoc}};
    \draw [nextptr] (WonOscars) -- (eoc);

    \node (AI) [head, minimum width=0, right of=ActIn, xshift=3.8cm, yshift=-1cm] {Act In};
    \draw [pntrlink] (ActIn) -| (AI);
    \node (IsCineTerm) [ndot, yshift=-0.3cm, below of=AI] {\tiny \texttt{0x4}};
    \draw [nextptr] (AI) -- (IsCineTerm);
    \node (IsA1) [pid1, xshift=-0.2cm, right of=IsCineTerm] {is a};
    \draw [proplink] (IsCineTerm) -- (IsA1);
    \node (CineTerm) [pid2, xshift=-0.6cm, left of=IsCineTerm, text width=1.7cm] {cinematic term};
    \draw [proplink] (IsCineTerm) -- (CineTerm);
    
    \draw [nextptr] (IsCineTerm) |- (eoc);

    \node (TF) [head, minimum width=0, left of=ThisFilm, xshift=-1.8cm, yshift=0.8cm] {This Film};
    \draw [pntrlink] (ThisFilm) |- (TF);
    
    \node (IsFilm) [ndot, below of=TF, yshift=0.1cm] {\tiny \texttt{0x6}};
    \draw [nextptr] (TF) -- (IsFilm);
    \node (IsA2) [pid1, right of=IsFilm] {is a};
    \draw [proplink] (IsFilm) -- (IsA2);
    \node (Film) [pid2, left of=IsFilm] {film};
    \draw [proplink] (IsFilm) -- (Film);

    \node (TitleSully) [ndot, yshift=-0.3cm, below of=IsFilm] {\tiny \texttt{0x7}};
    \draw [nextptr] (IsFilm) -- (TitleSully);
    \node (Title) [pid1, right of=TitleSully] {title};
    \draw [proplink] (TitleSully) -- (Title);
    \node (Sully) [pid2, left of=TitleSully] {``Sully''};
    \draw [proplink] (TitleSully) -- (Sully);

    \node (ProtSs) [ndot, yshift=-1cm, below of=TitleSully] {\tiny \texttt{0x8}};
    \draw [nextptr] (TitleSully) -- (ProtSs);
    \node (Prot) [pid1, xshift=0.5cm, right of=ProtSs] {protagonist};
    \draw [proplink] (ProtSs) -- (Prot);
    \node (Sully2) [pid2, left of=ProtSs] {Sully};
    \draw [proplink] (ProtSs) -- (Sully2);

    \draw [nextptr] (ProtSs) |- (eoc);
    
    \draw [fakejumper] (ThisFilm) to [bend right=30] (TF.north east);
    \draw [very thick, ->, densely dashed, draw=red] (ThisFilm) to [bend right=30] (TF.north east);
    \draw [very thick, ->, densely dashed, draw=red] (TF) to [bend right=30] (IsFilm);
    \draw [very thick, ->, densely dashed, draw=red] (IsFilm) to [bend right=30] (TitleSully);
    \draw [very thick, ->, densely dashed, draw=red] (TitleSully) to [bend right=30] (ProtSs);
    \draw [very thick, ->, densely dashed, rounded corners=10pt, draw=red] (ProtSs.south west) |- (eoc.south west);

    \node (F) [head, minimum width=0, below of=TF, xshift=-2.5cm, yshift=-5cm] {Film};
    \draw [pntrlink] (Film) -| (F);
    \node (IsForm) [ndot, below of=F] {\tiny \texttt{0xa}};
    \draw [nextptr] (F) -- (IsForm);
    \node (IsA3) [pid1, xshift=0.2cm, right of=IsForm] {is a};
    \draw [fakejumper] (IsForm) -- (IsA3);
    \draw [proplink] (IsForm) -- (IsA3);
    \node (Form) [pid2, xshift=0.2cm, left of=IsForm] {form};
    \draw [proplink] (IsForm) -- (Form);

    \draw [nextptr] (IsForm) |- (eoc);

    \node (SS) [head, minimum width=0, text width=2.25cm, below of=TH, xshift=3cm, yshift=-3.5cm] {Sully Sullenberger};
    \draw [fakejumper, <->, >=stealth] (Sully1) to [bend right=30] (SS.west);
    \draw [pntrlink] (Sully1) to [bend right=30] (SS.west);
    \draw [fakejumper] (Sully2) to [bend left=15] (SS);
    \draw [pntrlink] (Sully2) to [bend left=15] (SS);
    
    \node (IsFigure) [ndot, yshift=-0.5cm, below of=SS] {\tiny \texttt{0xc}};
    \draw [nextptr] (SS) -- (IsFigure);
    \node (IsA4) [pid1, right of=IsFigure] {is a};
    \draw [proplink] (IsFigure) -- (IsA4);
    \node (Figure) [pid2, text width=1.5cm, left of=IsFigure, xshift=-0.5cm] {public figure};
    \draw [proplink] (IsFigure) -- (Figure);

    \node (ProfPilot) [ndot, below of=IsFigure] {\tiny \texttt{0xd}};
    \draw [nextptr] (IsFigure) -- (ProfPilot);
    \node (Prof) [pid1, right of=ProfPilot, xshift=0.5cm] {profession};
    \draw [proplink] (ProfPilot) -- (Prof);
    \node (Pilot) [pid2, left of=ProfPilot, xshift=-0.5cm] {pilot};
    \draw [proplink] (ProfPilot) -- (Pilot);
    
    \draw [nextptr] (ProfPilot) |- (eoc);

  \end{tikzpicture}
  \caption{
    A \ac{gdb} example containing 5 chains: \textit{Tom Hanks}, \textit{Act In}, \textit{This Film}, \textit{Sully Sullenberger} and \textit{Film}, interconnected by pointers.
    Cyan arrows follow \texttt{next} pointers; black arrows represent primary or property pointers.
    Dashed red arrows indicate the traversal path to retrieve the contents from the \texttt{primID2} pointer at address \texttt{0x1}.
    The full contents of the chain \textit{Film} can be found in \Cref{fig:film_subs}.
  }
  \label{fig:th_gdb}
\end{figure*}
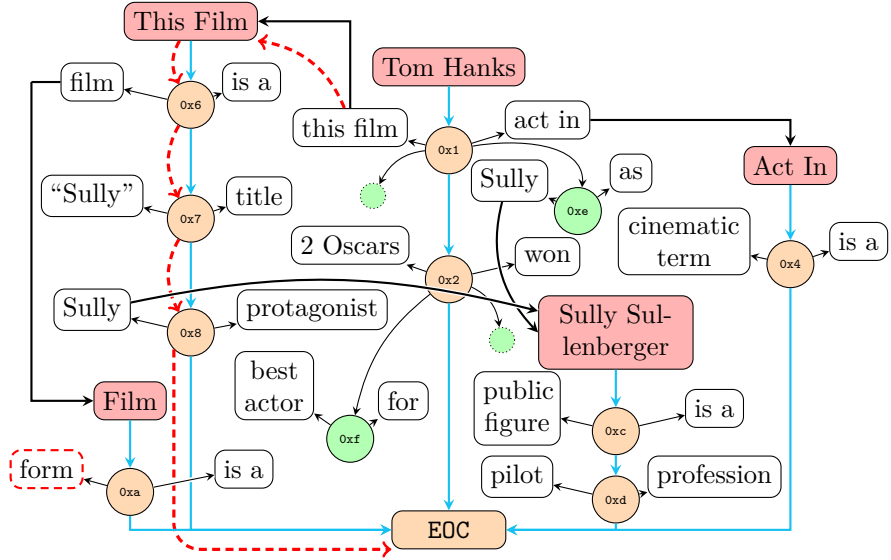

We are now ready to build a small \ac{gdb} using the \textit{Views} model and see its various features in action, as well as introduce some new details, as we shall see.
The example \ac{gdb} is illustrated in \Cref{fig:th_gdb}.
Note how in this example each linknode has been explicitly furnished with a physical address, visible inside the orange or green circle that represents it.

First let's note that the relationship encoding the phrase ``Tom Hanks - acts in - this film''  is covered by the linknode at address \texttt{0x1}, where \texttt{head ID} = ``Tom Hanks'', \texttt{primID1} = ``acts in'' and \texttt{primID2} = ``this film''.
Note that ``this film'' is just some generic, human-readable code-phrase we use to denote the chain that contains information on the film ``Sully''; we could have equally well used any other code-phrase such as ``linknode at \texttt{0x5}'' or ``entity 83''.
If we want to retrieve information about the film in question, we follow the pointer from \texttt{primID2}.
This leads to the ``This Film'' chain where we have stored three linknodes (\texttt{0x6}, \texttt{0x7} and \texttt{0x8}).
Following the retrieval path indicated by the dashed red arrows, they inform us that: (i) the entity is a film, (ii) the title is ``Sully'' and (iii) the protagonist is Sully.
Similarly, if we want to find out more about the general concept of ``act in'', we follow \texttt{primID1} towards \texttt{0x4}.

Next, what if we want to find \textit{contextual information}, such as: ``Who in this film does Tom Hanks act as \textit{specifically} (answer: the character, captain Sully Sullenberger)?'' 
Following \texttt{primID1} does not lead to an answer; it only returns general and non-contextual information about what it means to ``act in'' (\eg, it is a ``cinematic term'').
Thus, we create a subordinate chain and attach the green-colored linknode at \texttt{0xe}, specifying that ``act(s) in'' has the additional property ``as - Sully''; the same ``Sully'' that linknode \texttt{0x8} from the chain ``This Film'' is pointing to.
The main question left now is, why not add linknode \texttt{0xe} directly into the chain of ``Tom Hanks''?
Because ``as - Sully'' is a property of ``act(s) in'' within the context of ``This Film'', not of Tom Hanks in general.
Note: once again ``Sully'' is a human-readable codeword.
The key feature is that the corresponding \texttt{primID} from the linknode at \texttt{0xe} points to the ``Sully Sullenberger'' chain.

We continue exploring the information contained in the database.
The node at \texttt{0x7} informs us that the chain describes a film \textit{titled} ``Sully''. 
This is denoted explicitly as ``Sully'' to show that the pointer does not point to a linknode of the person Sully Sullenberger, but rather to a generic string.
This newly introduced feature is key: it allows the \ac{gdb} to refer to objects outside its direct space of linknodes and thus ``ground'' itself semantically.
In this manner, arbitrary objects such as strings, multimedia, and even ensemble activations of neural networks may be connected to the fabric of the \ac{gdb}.
Further study of this, however, lies outside the scope of this paper.

Next, let us consider the question: ``Who is Sully?'' 
If we can identify the chain with headnode ``Sully Sullenberger'' as the ``Sully''; in question, we can read the chain containing linknodes \texttt{0xc} and \texttt{0xd}, and find out that he is a public figure and a pilot by profession.
Note, however, that in this example reading the chain will not inform us that ``Sully'' is a protagonist in the film titled ``Sully''.
This is equivalent to being prompted with ``What do you know about Sully Sullenberger?'' and replying with the contents of the corresponding chain.
The information that he is a protagonist in the film titled ``Sully'' does not ``spring to mind''.
To obtain that information one needs a prompt of a sort like: ``In what film is Sully Sullenberger a protagonist''?
Now we have multiple cues: we know we are looking for an intersection of ``Sully Sullenberger'' and the concept of ``protagonist'', both of whose headnode addresses we assume to know (that is the meaning of being able to ``cue'' these concepts).
Critically, a content-addressable search in the database for where the cued concepts meet will allow us to recover the answer to our question, even though it lies in a linknode that does not exist in the chains of either of the cues!
(Instead it is to be found in a third chain, ``This Film''.)

Our next case will very briefly exemplify follow-up questions and knowledge build-up.
One may ask: ``Where does Tom Hanks act?'' and receive the answer ``He acts in Sully.''
Someone unfamiliar with the film may ask: ``What is Sully?'' and receive the answer ``It is a film.''
This allows the listener to start building the chain that we have denoted here as ``This film'' (in fact there is enough information to build the headnode and linknodes at \texttt{0x6} and \texttt{0x7}).
By someone not knowing what a film is, the follow-up question ``What is a film?'' can be also asked, and so on.

For our final example let us consider the following: 
Suppose we want to make a clear distinction between the real Sully Sullenberger and his on-screen character.
In this case we have two immediately obvious options: either (i) we will start a subordinate chain for the \texttt{primID} pointing to ``Sully'' in the linknode at \texttt{0x8}, or (ii) we will create a new chain for the dramatised version of Sully Sullenberger and let (or indeed ``rewire'') the connection from the \texttt{primID} of the linknode at \texttt{0x8} to this new chain, and then somewhere in the new chain introduce a linknode making it clear that the chain refers to the dramatis personae of the real Sully Sullenberger.
This illustrates both the importance of setting up a schema so as to be fit-for-purpose and the flexibility of the \textit{Views} data structure.
We note that the specific mechanics allowing ``rewiring'' as illustrated here bear a more than passing resemblance to the notion of \textit{schema learning} in psychology, but further exploration of this concept lies outside the scope of this paper.

We conclude the section with the remark that \textit{Views} is also naturally compatible with the property graph data model with ``nodes'' mapping to ``headnodes'', ``edges'' mapping to primIDs and ``properties'' possible to attach via subordinate chains.

\section{Materials and Methods: Hardware-Aware Implementation}
\label{sec:implementation}
The proposed model was designed for a regular mapping to hardware memory arrays and associative operations.
This gives the model a uniform organization for the search and traversal primitives evaluated below.
(See examples in \Cref{sec:results}.)
Furthermore, the structure of linknodes strongly prescribes implementations where each element of the linknode is stored in a separate memory array. 

This section first presents two mappings of the \textit{Views} linknode fields to physical memory arrays: the \textit{CNSM} and \textit{Normalized} allocations.
It then summarizes the ASOCA implementation lineage, identifies the \ac{asoca3} \ac{fpga} as the platform evaluated in \Cref{sec:khop}, and defines the associative operations used by the worked examples and traversal workload.

\subsection{Mapping linknodes to physical arrays: Two Allocations}
\label{sec:cnsm}
Eight functionally identical memory arrays are used in the \textbf{CNSM} allocation.
Each memory array is allocated an identifier reflecting its functional meaning as shown in \Cref{tab:cnsm}. 
``C'' arrays store primIDs, \ie, the main ``Content'' of the \textit{Views}-based \ac{gdb}. 
``N'' arrays store pointers that allow traversals, \ie, they are ``Navigational'' in nature. 
``S'' arrays store pointers that branch towards ``Subordinates''. 
Finally, ``M'' arrays were added to store extra properties (``Miscellaneous'') while bringing the total number of arrays to a power of 2.
This allocation is used in the worked examples and traversal mapping of \Cref{sec:results}.

\begin{table}[!htb]
  \caption{\textit{CNSM} allocation by array}
  \label{tab:cnsm}
\centering
\begin{tabular}{lllll}
  \toprule
  Type/function                  & Identifier  & Linknode mapping & Usage                            \\
  \midrule
  \multirow{2}{*}{Content}       & \textit{C1} & \texttt{primID1} & Edge vertex pointer              \\
                                 & \textit{C2} & \texttt{primID2} & Destination vertex pointer       \\
  \hline                                                          
  \multirow{2}{*}{Navigator}     & \textit{N1} & \texttt{head ID} & Source vertex pointer            \\
                                 & \textit{N2} & \texttt{next}    & Next linknode pointer            \\
  \hline                                                          
  \multirow{2}{*}{Subordinate}   & \textit{S1} & \texttt{prop1}   & Edge subordinate                 \\
                                 & \textit{S2} & \texttt{prop2}   & Destination subordinate          \\
  \hline                                                          
  \multirow{2}{*}{Miscellaneous} & \textit{M1} & \texttt{prop1}   & Edge universal properties        \\
                                 & \textit{M2} & \texttt{prop2}   & Destination universal properties \\
  \hline
\end{tabular}
\vspace*{-4pt}
\end{table}

We now turn to M arrays. 
They are introduced for convenience when translating the \textit{Views} data structure into hardware, allowing storage of fixed-format properties used throughout a schema, such as edge weights or vertex degrees.
For brevity, we will refer to these as ``\textit{universals}''.
The configuration of universals is an implementation-specific optimization for direct property retrieval and depends on the schema.
Notably, universals could be stored as additional linknodes, but their universal applicability means they can be essentially ``hard-wired'' within the M arrays.

Finally, we note that even schemas such as \textit{CNSM} can be further tweaked.
In this particular example we chose to designate C1 as ``edge pointer'' and C2 as ``destination vertex pointer'', clearly segregating edges from destination vertices.
This is not obligatory, however, as pointed out in the previous section.

\begin{table}[!htb]
  \caption{\textit{Normalized} allocation by array}
  \label{tab:normalised}
\centering
\begin{tabular}{lllll}
  \toprule
  Type                       & Identifier  & Linknode mapping & Usage                      \\
  \midrule
  \multirow{2}{*}{Content}   & \textit{C1} & \texttt{primID1} & Edge vertex pointer        \\
                             & \textit{C2} & \texttt{primID2} & Destination vertex pointer \\
  \hline                                                      
  \multirow{2}{*}{Navigator} & \textit{N1} & \texttt{head ID} & Source vertex pointer      \\
                             & \textit{N2} & \texttt{next}    & Next linknode pointer      \\
  \hline
\end{tabular}
\vspace*{-4pt}
\end{table}

Next, the \textbf{Normalized} allocation is a minimalistic version for simpler graphs, emphasizing compactness over complete functional flexibility.
As illustrated in \Cref{tab:normalised}, S and M arrays are removed, leaving just C and N behind.
This allocation may still represent subordinate chains by treating them as separate chains, but analysis of this possibility lies outside the scope of this paper.
The \textit{normalized} allocation is more suitable for graphs with less context-dependent information.
Finally, we note that S arrays and/or M arrays can be optionally supplemented upon the \textit{Normalized} allocation to produce further options for the \ac{gdb} designer. 

\subsection{ASOCA Lineage and Associative Interface}
\label{sec:asoca}

ASOCA comprises a sequence of implementations that map \textit{Views} fields to dually addressable memory and associative operations.
\ac{asoca1} established a $64\times64$-bit \ac{dam} array, built using beyond-CMOS memory cells, that stores a pointer field from either allocation \citep{strukov2008missing,pan2021rram,pan2024energy}.
\ac{asoca2} combined eight such arrays, one for each CNSM field, with local processing peripherals in a supercluster.
It provides architectural and pre-silicon evidence for the mapping.

The traversal measurements reported below were obtained from \ac{asoca3}, which implements the \textit{Views} execution model on an Alpha Data ADM-PCIE-9V7 board with a Xilinx VU13P \ac{fpga}.
It preserves the relevant CNSM fields and CAR/AAR-style operations, but iterates other architectural aspects of the system.
The \ac{asoca3} processing logic runs at 100~MHz and the board interface at 250~MHz.
The system contains 256 superclusters, each storing 512 linknodes.
It uses 82.03\% of the available logic look-up tables, 34.60\% of the registers and 39.86\% of the on-chip block memory.
This paper only reports the benchmark-relevant top-level facts in this commercially confidential implementation.

The following operations connect the \textit{Views} representation to the hardware.
They are used in the worked examples, but only CAR and AAR-family accesses are exercised directly in the K-hop evaluation; the host computer manages the frontier and visited set.

\begin{enumerate}
  \item Program (PROG): Sets a pointer within a linknode (e.g. the \texttt{primID1} pointer of some linknode N, stored in a C1-type array within a supercluster).
  \item \ac{aar}: Reads the pointer stored at a specified address in the database. 
  It acts as a standard memory read in conventional memories.
  \item \ac{car}: A pointer is searched for in the database as a cue. 
  The address(es) where it has been found are returned by the underlying associative memory.
  We can specify which array within a supercluster to search (e.g. in the \textit{CNSM} mapping: C1, C2, N1, ...).
  \item 2-sided content-addressable read (CAR2): An alternative version of \ac{car}, where we look for specific combinations of 2$\times$ pointers (for example we can look for specific combinations of pointers at arrays C1 and C2).
  \item A set of extra utility operations (HEAD, CARNEXT \& TAIL): Composite operations for traversing the graph structure.
  HEAD reads N1 of a given linknode and finds the headnode of the chain that ``owns'' this linknode. 
  CARNEXT returns ``the next match'' in the event that a CAR/CAR2 operation identifies more than a single match/answer.
  TAIL iteratively reads N2 until the \ac{eoc} and returns the address of the last linknode of the chain that `owns' a given linknode.
\end{enumerate}

The cost of these operations depends on how they are used and implemented.
Following N2 through a known finite chain is linear in the number of traversed linknodes.
CAR and CAR2 provide associative search, but returning multiple matches through CARNEXT still takes time in proportion to the number of matches.
Similarly, subordinate-chain retrieval is linear in the number of records visited and requires a depth limit or visited set when the data do not already impose a safe bound.
These properties do not imply constant-time graph traversal.

Here is a basic retrieval example: after programming a complete \textit{Views} image into the hardware with PROGs, \acp{car} on N1-type arrays can find the addresses of linknodes belonging to a specific headnode (\ie, they can ``highlight'' a complete chain).
This can be used directly for queries of the sort: ``Fetch all information \textit{directly} associated with Tom Hanks.''
Note the use of ``directly'' to denote that we are not looking for information about Tom Hanks available in \textit{other chains}, i.e. cases where thinking of a film or some other concept would ``make us think of'' Tom Hanks.

Afterwards, \acp{aar} can be used to retrieve further contents from the identified linknodes (e.g. the primIDs) and fuel further searches using \ac{car}, \ac{aar}, HEAD, CARNEXT or TAIL. 
Another example would be using CAR2 to locate 2 out of the 3 components in a ternary relationship, using a follow-up \ac{aar} to retrieve the final component. 
This can answer queries like: ``Who won 2 Oscars?'' (see \Cref{fig:th_gdb}).
We would send a CAR2 operation using ``won'' and ``2 Oscars'' as cues and then trace the head of the chain that owns nodes matching that description.

The \ac{isa} under discussion is organized around \textit{Views}, while associative search from hardware memory arrays provides a retrieval primitive for \textit{Views}-based graphs.
It illustrates the close alignment between the \textit{Views} data structure and the \ac{asoca} hardware architecture.
The same co-design methodology may be investigated for other graph kernels, but their behavior is not established by the operations defined here \citep{dijkstra2022note,blondel2008fast}.

\section{Results}
\label{sec:results}
This section first characterizes storage footprint across the evaluated social-network snapshots and then reports directed K-hop traversal on the \ac{asoca3} \ac{fpga} implementation.
The later symbolic-retrieval and Copycat-inspired examples are representational case studies rather than quantitative workload benchmarks.

\subsection{\acs{ldbc} Storage Footprint}
\label{sec:storage}
The storage study uses the \acf{ldbc} \acf{snb} Interactive v1 Datagen v1.0.0 initial snapshots \citep{erling2015ldbc,angles2020ldbc} at three \acp{sf}: 0.1, 0.3 and 1.
It measures storage only; it does not execute the Interactive workload.
\acs{ldbc} \acs{snb} was selected because it provides a recognized, reproducibly generated social graph at controlled scales.
Its labels, repeated properties, multiple relation types, parallel edges and relation-local edge properties exercise the mapping choices considered here.
At \acs{sf}1, the canonical inventory contains 3,181,724 entities, 17,256,038 identified edges, 19,011,685 node-property occurrences and 4,012,190 edge-property occurrences.
It maps to 58,707,660 valid \textit{Views} rows; these counts describe the largest evaluated snapshot rather than the structural variety or scale of knowledge graphs in general.
The checksum-pinned \texttt{CsvBasic} snapshots with \texttt{LongDateFormatter} are converted into a canonical stream that retains entities, labels, repeated node properties, identified and parallel edges, edge properties, literal datatypes and source order.
The corresponding publisher Turtle archives are validated separately and are not treated as statement-for-statement equivalents because their mapping omits or restructures some \acf{csv} content.

In the \textit{Views} mapping, entities, vocabulary terms and distinct typed literals are headnodes; node labels, node properties and edges are main-chain linknodes; and edge properties retain relation-local scope.
The CNSM allocation uses an S1 subordinate chain located by \ac{aar} from its parent edge row, while the Normalized allocation stores a separately discoverable property chain whose N1 field identifies the parent edge row and is located by \ac{car}.
The latter preserves parallel-edge identity and relation-local scope, but exchanges a direct pointer read for an associative lookup rather than discarding edge properties.
Address zero is reserved for \texttt{NULL}/\ac{eoc}.
Four validated little-endian images vary the allocation (four-array Normalized or eight-array CNSM) and entry width (32 or 64 bits).
Including the reserved zero address, the largest image requires 26 address bits.
Thus, 32-bit entries are sufficient for these local row addresses, while the 64-bit images expose width sensitivity and provide direct comparators for the 64-bit identifier slots used in the alternative representations.

The analytical accounting separates logical structures from native database record formats.
Each CNSM row contains eight fixed-width slots and each Normalized row contains four.
The \ac{lpg} mapping uses two 64-bit slots per node--label record, four per identified edge and three per node- or edge-property occurrence.
The lossless \ac{rdf} mapping uses one triple per label or node property, two triples to represent each identified edge as a resource, and one additional triple per edge property; each triple contains three 64-bit slots.
The same deterministic vocabulary and typed-literal dictionaries are added to every analytical total.

\begin{table}[!htb]
  \caption{Analytical storage for \acs{ldbc} \acs{snb} Interactive v1 initial snapshots (MiB, including shared dictionaries). LPG and RDF use the stated logical 64-bit slot accounting rather than native database layouts.}
  \label{tab:ldbc-analytical}
  \centering
  \resizebox{\textwidth}{!}{%
  \begin{tabular}{lrrrrr}
    \toprule
    \acs{sf} & 32-bit Normalized & 64-bit Normalized / 32-bit CNSM & 64-bit CNSM & LPG & RDF \\
    \midrule
    0.1 & 176.25 & 258.42 & 422.76 & 192.94 & 218.07 \\
    0.3 & 521.79 & 765.41 & 1252.65 & 575.85 & 652.78 \\
    1 & 1928.79 & 2824.60 & 4616.21 & 2135.26 & 2422.91 \\
    \bottomrule
  \end{tabular}%
  }
\end{table}

\begin{figure}[!htb]
  \centering
  \begin{tikzpicture}[x=5.0cm,y=0.85cm]
    \draw[->] (0,0) -- (1.08,0) node[right] {\acs{sf}};
    \draw[->] (0,0) -- (0,5.35) node[above] {Total (GiB)};
    \foreach \x/\xlabel in {0.1/0.1,0.3/0.3,1/1} {
      \draw (\x,0.06) -- (\x,-0.06) node[below] {\xlabel};
    }
    \foreach \y in {1,...,5} {
      \draw (0.008,\y) -- (-0.008,\y) node[left] {\y};
    }
    \draw[blue,dashed,mark=triangle*] plot coordinates {(0.1,0.172120) (0.3,0.509560) (1,1.883582)};
    \draw[cyan!65!black,thick,mark=o] plot coordinates {(0.1,0.252365) (0.3,0.747470) (1,2.758395)};
    \draw[blue,thick,mark=*] plot coordinates {(0.1,0.412855) (0.3,1.223292) (1,4.508019)};
    \draw[red,thick,mark=x] plot coordinates {(0.1,0.188420) (0.3,0.562353) (1,2.085215)};
    \draw[green!55!black,thick,mark=+] plot coordinates {(0.1,0.212955) (0.3,0.637484) (1,2.366126)};
    \begin{scope}[shift={(0.04,5.02)}]
      \draw[blue,dashed,mark=triangle*] (0,-0.00) -- (0.04,-0.00) node[right,font=\scriptsize,text=black] {32-bit Normalized};
      \draw[cyan!65!black,thick,mark=o] (0,-0.32) -- (0.04,-0.32) node[right,font=\scriptsize,text=black] {64-bit Normalized / 32-bit CNSM};
      \draw[blue,thick,mark=*] (0,-0.64) -- (0.04,-0.64) node[right,font=\scriptsize,text=black] {64-bit CNSM};
      \draw[red,thick,mark=x] (0,-0.96) -- (0.04,-0.96) node[right,font=\scriptsize,text=black] {LPG};
      \draw[green!55!black,thick,mark=+] (0,-1.28) -- (0.04,-1.28) node[right,font=\scriptsize,text=black] {RDF};
    \end{scope}
  \end{tikzpicture}
  \hfill
  \begin{tikzpicture}[x=0.85cm,y=0.85cm]
    \draw[->] (-0.65,0) -- (5.2,0);
    \draw[->] (-0.55,0) -- (-0.55,5.35) node[above] {\acs{sf}1 (GiB)};
    \fill[blue!55] (-0.22,0) rectangle (0.22,0.874812);
    \fill[gray!65] (-0.22,0.874812) rectangle (0.22,1.883582);
    \node[rotate=55,anchor=east,font=\scriptsize] at (0,-0.10) {32-bit Normalized};
    \fill[blue!55] (0.78,0) rectangle (1.22,1.749625);
    \fill[gray!65] (0.78,1.749625) rectangle (1.22,2.758395);
    \node[rotate=55,anchor=east,font=\scriptsize] at (1,-0.10) {\shortstack{64-bit Normalized\\32-bit CNSM}};
    \fill[blue!55] (1.78,0) rectangle (2.22,3.499249);
    \fill[gray!65] (1.78,3.499249) rectangle (2.22,4.508019);
    \node[rotate=55,anchor=east,font=\scriptsize] at (2,-0.10) {64-bit CNSM};
    \fill[blue!55] (2.78,0) rectangle (3.22,1.076445);
    \fill[gray!65] (2.78,1.076445) rectangle (3.22,2.085215);
    \node[rotate=55,anchor=east,font=\scriptsize] at (3,-0.10) {LPG};
    \fill[blue!55] (3.78,0) rectangle (4.22,1.357356);
    \fill[gray!65] (3.78,1.357356) rectangle (4.22,2.366126);
    \node[rotate=55,anchor=east,font=\scriptsize] at (4,-0.10) {RDF};
    \foreach \y in {1,...,5} {
      \draw (-0.62,\y) -- (-0.48,\y) node[left] {\y};
    }
    \fill[blue!55] (3.55,4.95) rectangle (3.85,5.15) node[right,xshift=2pt,font=\scriptsize,text=black] {structure};
    \fill[gray!65] (3.55,4.55) rectangle (3.85,4.75) node[right,xshift=2pt,font=\scriptsize,text=black] {dictionary};
  \end{tikzpicture}
  \caption{Observed analytical storage over the three evaluated \acs{snb} scale factors (left) and the \acs{sf}1 decomposition into structural and shared-dictionary bytes (right). Lines connect the evaluated snapshots and do not establish behavior beyond \acs{sf}1.}
  \label{fig:ldbc-storage-growth}
\end{figure}

\Cref{tab:ldbc-analytical,fig:ldbc-storage-growth} show that field count, entry width and dictionary storage all contribute materially to the totals.
The 64-bit Normalized and 32-bit CNSM allocations have the same 32 structural bytes per row, which isolates array count from entry width.
For this mapping, S2, M1 and M2 remain reserved and zero in the CNSM images, while S1 is used only by property-bearing edges.
Consequently, 64-bit CNSM is 116--119\% larger than the logical LPG totals and 91--94\% larger than the logical RDF totals at the three scales.
Conversely, 32-bit Normalized is 9--10\% smaller than LPG and 19--20\% smaller than RDF under the stated accounting, at the cost of associative subordinate-chain discovery.
These results demonstrate allocation and width specialization rather than an intrinsic storage advantage of \textit{Views} over the 64-bit alternatives.

For the materialized comparison, the same canonical graph is written as four self-contained \textit{Views} images and loaded into Neo4j Community 2026.01.4 and Apache Jena TDB2 5.0.0.
Neo4j receives typed bulk-import \acs{csv} with no requested user indexes; the isolated database is count-checked, shut down cleanly and measured separately from its transaction logs.
TDB2 receives validated deterministic N-Triples in which identified edges are resources, preserving parallel-edge identity and edge properties; the complete closed, uncompacted location is measured.
Input exports, reports, process logs, the Neo4j system database and temporary files are excluded.

\begin{table}[!htb]
  \caption{Materialized post-mapping and post-load footprint (MiB). App. and Alloc. denote apparent and allocated bytes.}
  \label{tab:ldbc-materialized}
  \centering
  \resizebox{\textwidth}{!}{%
  \begin{tabular}{lrrrrrr}
    \toprule
     & \multicolumn{2}{c}{\acs{sf}0.1} & \multicolumn{2}{c}{\acs{sf}0.3} & \multicolumn{2}{c}{\acs{sf}1} \\
    Representation & App. & Alloc. & App. & Alloc. & App. & Alloc. \\
    \midrule
    Views: 32-bit Normalized & 176.26 & 176.27 & 521.79 & 521.81 & 1928.79 & 1928.82 \\
    Views: 64-bit Normalized & 258.43 & 258.44 & 765.41 & 765.45 & 2824.60 & 2824.64 \\
    Views: 32-bit CNSM & 258.43 & 258.44 & 765.41 & 765.45 & 2824.60 & 2824.65 \\
    Views: 64-bit CNSM & 422.77 & 422.79 & 1252.66 & 1252.71 & 4616.22 & 4616.28 \\
    \midrule
    Neo4j database & 422.76 & 422.80 & 1137.02 & 1137.06 & 4076.33 & 4076.37 \\
    Neo4j transaction logs & 257.00 & 257.00 & 257.00 & 257.00 & 257.00 & 257.01 \\
    Jena TDB2 location & 994.61 & 822.24 & 2658.77 & 2488.38 & 9376.08 & 9218.69 \\
    \bottomrule
  \end{tabular}%
  }
\end{table}

\Cref{tab:ldbc-materialized} reports each complete \textit{Views} allocation directory, the Neo4j target database and its transaction logs separately, and the complete closed TDB2 location.
This materialized comparison answers a different question from the analytical accounting: it measures deployable array images and native software stores after lossless, model-appropriate translations of the same canonical records.
The larger TDB2 footprint is partly conditioned by the edge-resource representation used to preserve parallel-edge identity and edge properties, while the Neo4j footprint is conditioned by native relationship records and stored canonical identifiers.

At \acs{sf}0.1, the \textit{Views} images and native stores were materialized twice; the Views counts and content hashes were identical, the native inventories matched, and all allocated-footprint ranges were below 0.005\%.
The pre-specified 1\% trigger for a third repetition was therefore not reached, and \acs{sf}0.3 and \acs{sf}1 were materialized once under the same validation procedure.
The deterministic values in \Cref{tab:ldbc-analytical,tab:ldbc-materialized} are reported directly without inferential testing.
They show the effects of field count, entry width, dictionaries and native-store facilities under the stated mappings.
Unlike the software stores, the \textit{Views} images do not contain a transaction manager, query planner, page cache or general-purpose indexes.
The materialized figures therefore establish post-mapping and post-load footprints, not an intrinsic ordering of the underlying models.
Neither comparison provides evidence about query latency, throughput or runtime scalability.

\subsection{K-hop Traversal on the ASOCA3 FPGA}
\label{sec:khop}
The traversal workload returns the distinct vertices reachable from a selected source by at most $k$ directed outgoing edges.
The host computer tracks the vertices to explore next and those already visited.
It removes duplicates, does not preserve multiple paths to the same vertex and does not revisit a vertex.
Execution terminates when $k$ frontiers have been processed or the frontier becomes empty.
For each frontier vertex, the host uses CAR over N1 to identify source-owned relations and AAR-family reads over C2 to retrieve their destinations from the resident graph.
The experiment therefore measures a traversal shared between the host computer and the accelerator, rather than a wholly on-chip implementation.

The input contains the first 131,072 recorded directed relationships of the SNAP roadNet-CA dataset (\url{https://snap.stanford.edu/data/roadNet-CA.html}), comprising 45,915 vertices with mean out-degree 2.89 and recorded range 1--8.
The retained workload uses the same 1,000 fixed source vertices for $k\in\{2,5,8,10,15,20\}$.
Each source was timed once at each hop bound, giving 6,000 timed cases.
The case record retains the number of vertices first discovered at every hop and the cumulative number of distinct vertices returned through $k$.
Its two representations have been checked for identical sources, frontier counts and totals.
The exact subset extraction, graph encoding and loading procedure were not retained, so the measurements remain descriptive evidence for this fixed graph image.

The retained benchmark reports two timing boundaries.
``Core'' is the time attributed to the accelerator, while ``system'' also includes communication with the host computer.

\begin{table}[!htb]
  \caption{Descriptive K-hop results across 1,000 fixed sources per hop bound. Core and system means use the separately reported timing boundaries.}
  \label{tab:khop-latency}
  \centering
  \resizebox{\textwidth}{!}{%
  \begin{tabular}{crrrrrrr}
    \toprule
    Hops & Cases & Mean returned vertices & \multicolumn{4}{c}{Core latency ($\mu$s)} & System mean (ms) \\
     & & & Mean & Std. & Min. & Max. & \\
    \midrule
    2 & 1,000 & 8.934 & 15.74 & 12.67 & 1.75 & 90.07 & 0.08825 \\
    5 & 1,000 & 48.275 & 83.12 & 54.49 & 9.72 & 440.37 & 0.46765 \\
    8 & 1,000 & 136.412 & 212.75 & 121.02 & 16.84 & 907.11 & 1.19 \\
    10 & 1,000 & 232.313 & 342.12 & 197.00 & 23.42 & 1520.50 & 1.93 \\
    15 & 1,000 & 648.624 & 888.89 & 520.43 & 47.62 & 3479.69 & 5.01 \\
    20 & 1,000 & 1377.963 & 1805.65 & 1202.88 & 65.07 & 10852.90 & 10.17 \\
    \bottomrule
  \end{tabular}%
  }
\end{table}

The mean number of returned vertices rises from 8.934 at $k=2$ to 1377.963 at $k=20$, while mean latency increases at both reported boundaries.
This observation characterizes increasing traversal work over one fixed graph image; it does not establish scaling with graph size or constant cost per hop.
Cost also depends on frontier size, graph structure, the number of matches and host-side bookkeeping.
The raw per-query latency observations and system-latency dispersion were not retained.
The reported standard deviations therefore describe variation across source queries, not repeated-run variability, and no confidence intervals or significance tests are reported.
Throughput, workload power, energy and per-access memory efficiency were not measured, and no controlled CPU/GPU comparison or general speed-up is claimed.
K-hop evaluates a graph-processing component relevant to multi-hop retrieval, but it does not constitute an end-to-end question-answering evaluation.

\subsection{Illustrative Symbolic Retrieval Composition}
\label{sec:reasoning}
This section presents one schema-dependent example of composing retrieval operations over explicit symbolic relations.
It illustrates a representational procedure, not a logic calculus or reasoning benchmark.
Inference accuracy, rule coverage and conflicting knowledge are not evaluated.
The example uses a simple syllogism in natural language to derive that ```This' (i.e. the object of discourse) is feline'':

\textbf{Major Premise:} `This' is a cat;

\textbf{Minor Premise:} Cats are feline;

\textbf{Conclusion:} \ \ \ \ \ \ `This' is feline.
  
\begin{figure*}[!htb]
\centering
\subfloat[]{
  \begin{tikzpicture}
    \node (0x00a) [head] {\texttt{0x00a}};
    \node (temper) [next, fill=orange!30, right of=0x00a, xshift=2cm] {temper-naughty};
    \node (color) [next, fill=orange!30, right of=temper, xshift=2.5cm] {color-black};
    \node (species) [next, fill=orange!30, right of=color, xshift=2cm, draw=blue, densely dashed, very thick] {species-cat};
    \node (eoc) [next, fill=orange!30, right of=species, xshift=1.5cm] {\ac{eoc}};

    \draw [nextptr] (0x00a) -- (temper);
    \draw [nextptr] (temper) -- (color);
    \draw [nextptr] (color) -- (species);
    \draw [nextptr] (species) -- (eoc);
    \node (black) [head, below of=0x00a, xshift=1.5cm] {Black};
    \node (lambda) [next, fill=orange!30, right of=black, xshift=1.5cm] {$\lambda$-none};
    \node (rgb) [next, fill=orange!30, right of=lambda, xshift=2cm] {RGB-\texttt{\#000000}};

    \draw [nextptr] (black) -- (lambda);
    \draw [nextptr] (lambda) -- (rgb);
    \draw [nextptr] (rgb) -| (eoc);

    \node (color2black) [coordinate, below of=color, yshift=0.5cm] {};
    \draw [thick, draw=black] (color) -- (color2black);
    \draw [pntrlink] (color2black) -| (black);
    \node (cat) [head, below of=black, xshift=1.5cm] {Cat};
    \node (persian) [next, fill=orange!30, right of=cat, xshift=2cm] {$\exists$breed-Persian};
    \node (family) [next, fill=orange!30, right of=persian, xshift=2.5cm, draw=red, densely dashed, very thick] {family-Felidae};

    \draw [nextptr] (cat) -- (persian);
    \draw [nextptr] (persian) -- (family);
    \draw [nextptr] (family) -| (eoc);
    
    \node (species2cat) [coordinate, below of=species, yshift=-0.5cm] {};
    \draw [fakejumper] (species) -- (species2cat);
    \draw [thick, draw=black] (species) -- (species2cat);
    \draw [pntrlink] (species2cat) -| (cat);
  \end{tikzpicture}
  \label{fig:eg_db}
}
\vfil
\subfloat[]{
  \begin{tikzpicture}[scale=0.9, transform shape]
    \node (triplet) [address] {\texttt{0x6fd}};
    \node (head) [head, above of=triplet, yshift=0.75cm] {Cat};
    \node (pid1) [pointer, right of=triplet, xshift=1cm, yshift=0.75cm] {family};
    \node (pid2) [pointer, left of=triplet, xshift=-1cm, yshift=0.75cm] {Felidae};
    \node (prop1) [prop, right of=triplet, xshift=1cm, yshift=-1cm, text width=1.5cm] {NULL};
    \node (prop2) [prop, left of=triplet, xshift=-1cm, yshift=-1cm, text width=1.5cm] {NULL};
    \node (next) [next, below of=triplet, yshift=-0.5cm] {\ac{eoc}};
    
    \draw [pntrlink] (triplet) -- (head);
    \draw [proplink] (triplet) to [out=60, in=180] (pid1);
    \draw [proplink] (triplet) to [out=120, in=0] (pid2);
    \draw [proplink] (triplet) to [bend left=30] (prop1);
    \draw [proplink] (triplet) to [bend right=30] (prop2);
    \draw [pntrlink] (triplet) -- (next);
  \end{tikzpicture}
  \label{fig:family_feline}
}
\caption{
  Part of the example knowledge base contents:
  (a) Chains of \texttt{0x00a} (from \Cref{fig:chain_sentence}) \textit{Black} and \textit{Cat} in \textit{Views} format. 
  (b) The \textit{family-Felidae} linknode (shown above in red, dashed lines in the \textit{Cat} chain), denoting the link among \textit{cat}, \textit{Felidae} and \textit{family}. 
}
\label{fig:cat_is_feline}
\end{figure*}
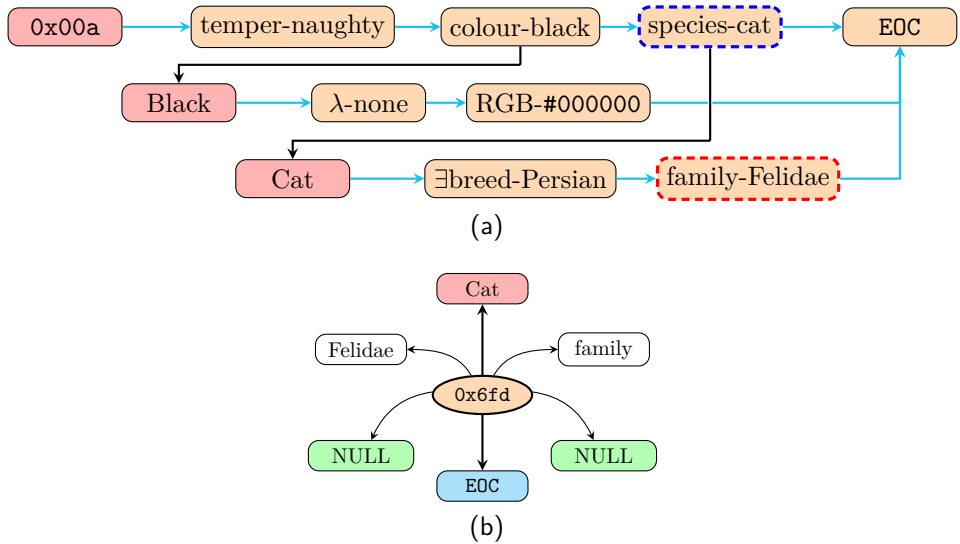

We represent the premises and compose retrieval calls in a small program that derives the stated conclusion under the selected schema.
For the representation let us choose a straightforward scheme:
First, a chain to represent the object we are attempting to make an inference about (``This'') including a linknode with primIDs ``species'' and ``cat'' (just as we have done in \Cref{fig:chain_sentence}).
Then, a chain for the general concept of a ``cat'' featuring a linknode whose primIDs are the pair (``family'', ``Felidae'') plus any other relevant chains.
These are illustrated in \Cref{fig:eg_db}.

Next, comes the program.
This code can run on a host machine with access to a \textit{Views} store.
We focus on the retrieval calls before presenting the full pseudocode in \Cref{alg:felidae_search}.
The calls use Python-like notation for simplicity.
We also omit trivial checks such as verifying that our CAR/AAR operations return valid results to focus on the logic of the algorithm rather than the minutiae.

The example begins by searching for the family of `this' using two CAR2 operations:

\begin{verbatim}
    results_C1 = CAR2(N1="this", C1="family")
    results_C2 = CAR2(N1="this", C2="family")
    results = [results_C1, results_C2]
\end{verbatim}
where our notation assumes the \textit{CNSM} configuration.
The main code can then check if the \verb|results| list contains the (``family'', ``Felidae'') pairing.
If present, this pairing gives the result directly.
The constructed example does not contain that direct pairing, so a second retrieval stage is required.

In the second stage, we will attempt to find the species of `this' and then see if we can infer the family from the species.
We therefore perform the following operations:

\begin{verbatim}
    addresses_1 = CAR2(N1="this", C1="species")
    addresses_2 = CAR2(N1="this", C2="species")
    results_C2 = AAR(addresses_1, C2)
    results_C1 = AAR(addresses_2, C1)
    results = [results_C1, results_C2]
\end{verbatim}

The CAR2 pair informs us whether `this' is in any way related to ``species'' and stores the addresses of the corresponding linknodes, from which we again perform AARs to tell us how `this' relates to ``species''.
For each result in \verb|results| (here the headnode of ``Cat''), we test for the pair (``family'', ``Felidae'') to answer the question: ``Is the species under discussion a member of the family of Felidae?''

\begin{verbatim}
    results_C1 = CAR2(N1=results, C1="family")
    results_C2 = CAR2(N1=results, C2="family")
    results = [results_C1, results_C2]
\end{verbatim}

Finding this pairing returns an affirmative result.
Other retrieval sequences are possible under the same schema.

The full pseudocode is shown in \Cref{alg:felidae_search}, with \textit{Views} retrieval operations highlighted in red.
Note how we compact the pairs on C1/C2 into a single pseudocode function for brevity.

Different representation schemes change the retrieval sequence.
For example, ``The species of `this' is cat'' could instead use the primID pair (``is'', ``cat'').
A hierarchical taxonomy might also place ``family'' beneath ``genus''.
The schema must therefore be selected with the intended retrieval procedure in mind.

\newcommand{\CAR}{\textcolor{red}{CAR}}
\newcommand{\CARtwo}{\textcolor{red}{CAR2}}
\newcommand{\AAR}{\textcolor{red}{AAR}}
\algrenewcommand\algorithmicrequire{\textbf{Input:}}
\algrenewcommand\algorithmicensure{\textbf{Output:}}

\begin{algorithm}[H]
\caption{Search for ``Felidae'' in `this' chain and its species chain.}
\label{alg:felidae_search}
\begin{algorithmic}[1]
  \Require{$\texttt{0x00a}, \texttt{Felidae}$}
  \Ensure{$felidaeAddr$}

  \For{$thisAddr$ \textbf{in} $\CARtwo(N1=\texttt{0x00a}, C1/C2=\texttt{family})$}
    \State $thisResult \gets \AAR(thisAddr, C1/C2)$
    \If{$\AAR(thisAddr, C2/C1) = \texttt{Felidae}$}
      \State \Return $thisAddr$ \Comment{Felidae found in `this' chain}
    \EndIf
  \EndFor
  \For{$thisAddr$ \textbf{in} $\CARtwo(N1=\texttt{0x00a}, C1/C2=\texttt{species})$}
    \State $thisResult \gets \AAR(thisAddr, C2/C1)$ \Comment{Species of `this'}
    \For{$speciesAddr$ \textbf{in} $\CARtwo(N1=thisResult, C1/C2=\texttt{family})$}
      \State $speciesResult \gets \AAR(speciesAddr, C1/C2)$
      \If{$\AAR(speciesAddr, C2/C1) = \texttt{Felidae}$}
        \State \Return $speciesAddr$ \Comment{Felidae found in its species chain}
      \EndIf
    \EndFor
  \EndFor
  \State \Return $NULL$ \Comment{Felidae NOT found}
\end{algorithmic}
\end{algorithm}

\subsection{Copycat-Inspired Representational Mapping}
Copycat is a cognitive model for string analogy problems such as ``if $abc \sim abz$, then $zyx \sim ?$'' or ``$abc:abz::zyx:?$'' \citep{hofstadter1995fluid,mitchell1993analogy}.
The model uses a static concept storage structure, organized as a graph consisting of vertices representing ``crisp'' (i.e. not probabilistic or otherwise ``fuzzy'') concepts.
This was called the \textbf{slipnet}. 

This example maps selected slipnet structures and activation-like quantities into \textit{Views}.
It does not execute or validate the Copycat model.

The name ``slipnet'' comes from the key operating principle of Copycat, the \textbf{slippage}.
It is a mechanism whereby a concept can be dynamically substituted by another during problem-solving \citep{french1996subtlety}.
For example, in the string analogy example above, Copycat may test the hypothesis that [\textit{3rd-letter-in-string} is \textit{last-letter-of-alphabet}].
An alternative hypothesis is that [\textit{3rd-letter-in-string} is \textit{first-letter-of-alphabet}].
Copycat explores it through slippage, replacing ``Last'' with ``First'' through the ``Opposite'' edge in \Cref{fig:slippage}.
As such, slippage allows Copycat to flexibly explore alternative hypotheses, even when concepts appear semantically inconsistent or contradictory, \eg, \textit{First-Last} or \textit{Leftmost-Rightmost}.

\begin{figure}[!htb]
  \centering
  \begin{tikzpicture}[scale=0.8, transform shape]
    \node (0xcafe) [address] {\texttt{0xcafe}};
    \node (cafeHead) [head, above of=0xcafe, yshift=0.75cm] {\texttt{First}};
    \node (cafePrimid1) [pointer, right of=0xcafe, xshift=1cm, yshift=0.75cm] {\texttt{opposite}};
    \node (cafePrimid2) [pointer, left of=0xcafe, xshift=-1cm, yshift=0.75cm] {\texttt{last}};
    \node (cafeProp1) [prop, right of=0xcafe, xshift=1.75cm, yshift=-0.75cm, text width=2cm] {\texttt{slip lock\\=\textcolor{blue}{True}}};
    \node (cafeProp2) [prop, left of=0xcafe, xshift=-1.75cm, yshift=-0.75cm, text width=2cm] {\texttt{slip lock\\=\textcolor{red}{False}}};
    \node (cafeNext) [next, below of=0xcafe, yshift=-1.25cm] {\texttt{0xcaff}};
    
    \draw [pntrlink] (0xcafe) -- (cafeHead);
    \draw [proplink] (0xcafe) to [bend left=30] (cafePrimid1);
    \draw [proplink] (0xcafe) to [bend right=30] (cafePrimid2);
    \draw [proplink, densely dashed] (0xcafe) to [bend left=30] (cafeProp1);
    \draw [proplink, densely dashed] (0xcafe) to [bend right=30] (cafeProp2);
    \draw [pntrlink] (0xcafe) -- (cafeNext);

    \node (Last) [head, left of=cafePrimid2, xshift=-3cm] {\texttt{Last}};
    \draw [pntrlink] (cafePrimid2) -- (Last);
    \node (0xc0c0) [address, below of=Last] {\texttt{0xc0c0}};
    \node (c0c0Prop1) [prop, right of=0xc0c0, xshift=1cm, yshift=-2cm, text width=4cm] {\texttt{Activ=30<80, Activ lock=\textcolor{blue!70}{True}}};
    \node (c0c0Omit) [below of=0xc0c0, fill=cyan!30] {...};
    
    \draw [pntrlink] (0xc0c0) -- (Last);
    \draw [proplink, densely dashed] (0xc0c0) to [bend left=30] (c0c0Prop1);
    \draw [pntrlink] (0xc0c0) -- (c0c0Omit);
    
    \node (Opposite) [head, right of=cafePrimid1, xshift=2.5cm] {\texttt{Opposite}};
    \draw [pntrlink] (cafePrimid1) -- (Opposite);
    \node (0xface) [address, below of=Opposite] {\texttt{0xface}};
    \node (faceProp1) [prop, thick, densely dashed, draw=red, right of=0xface, xshift=1cm, yshift=-2cm, text width=4cm] {\texttt{Activ=\textcolor{blue!70}{100}>80, Activ lock=\textcolor{red}{False}}};
    \node (faceOmit) [below of=0xface, fill=cyan!30] {...};

    \draw [pntrlink] (0xface) -- (Opposite);
    \draw [proplink, densely dashed] (0xface) to [bend left=30] (faceProp1);
    \draw [pntrlink] (0xface) -- (faceOmit);
    
    \draw [proplink, very thick, rounded corners=10pt, draw=blue!70] (cafeHead)
      -| node [midway, above=3pt, yshift=-0.15cm] {\textcolor{blue!70}{\textit{\textbf{Activ propagating}}}}
      (faceProp1);

    \draw [fakejumper] (faceProp1) to [bend left=5] (cafeProp2);
    \draw [proplink, very thick, draw=red] (faceProp1) to [bend left=5] (cafeProp2);
    \draw [proplink, very thick, draw=red] (cafeProp2) -- (Last);
    \draw [proplink, very thick, draw=red] (Last)
      -- node [midway, above] {\textcolor{red}{\textit{\textbf{Slip!}}}}
      (cafeHead);

    \node (activComment) [below of=c0c0Prop1, yshift=0.25cm, xshift=0.25cm, text width=5cm] {\textit{(Slipnet threshold is \textcolor{blue!70}{\texttt{80}})}};
  \end{tikzpicture}
  \caption{
    Example of slippage from \textit{Last} to \textit{First} via \textit{Opposite} in \textit{Views} slipnet. 
    The example database contains a very simple proposition, saying ``First is the opposite of last'' at address \texttt{0xcafe}.
    Note the purple rectangles representing the contents of the M arrays. 
    The slippage process begins with the activation propagating from \textit{First} to \textit{Opposite}, allowed because Activ lock is false (blue path).
    As a result, the activation value of \textit{Opposite} crosses the threshold (80), which then triggers the slippage to \textit{First} via \textit{Opposite}, after a slip-lock check conducted at \textit{Last} (\texttt{0xc0c0}).
  }
  \label{fig:slippage}
\end{figure}

Here, we detail a \textit{Views} representation of the slipnet and an illustrative slippage update.
\Cref{tab:slipnet} summarizes slipnet's data organization converted into the \textit{Views} \textit{CNSM} allocation.
This conversion of slipnet starts by storing each concept (known as \textbf{slipnodes} in Copycat's terminology) as a headnode in the N1 array (e.g. ``First''). 
Once slipnodes are instantiated, conceptual relationships are encoded using the regular (non-head) linknode format: C1 holds the primID pointer to the edge label, and C2 holds another to the destination concept. 
This encodes \textbf{sliplinks}, such as ``First -- Opposite -- Last'', within the \textit{Views} framework.
Note that some edges in Copycat are not explicitly labeled, in which case we assign said edges ad hoc \acp{id} pointing to headnodes without further linknodes in their chains.

Copycat is also an ``activation-based'' cognitive model and as with other such models \citep{collins1975spreading, anderson1996working, altmann2002memory, nuxoll2004comprehensive}, it works by propagating and periodically updating the levels of activation across the slipnet following predefined operation rules \citep{hofstadter1995fluid}.
This mechanism affects how operators in this model seek both regular and creative (slipped) connections, mirroring the cognitive processes underlying human creativity.
In order to operate Copycat-like activation-based mechanics under \textit{Views} we need to allocate space for storing: a) the activation value/strength ``Activ'' of each linknode, b) a ``natural activation decay'' value regulating how quickly activation decays in the absence of other updates (in original Copycat: ``conceptual depth''; for details please see original publication \citep{hofstadter1995fluid}) and c) a ``conductance'' value encoding how efficiently activation can spread over a linknode.
We also add, of our own accord, a couple of 1-bit flags for improving the operational flexibility of the system: the slip lock which enables/disables slippage along a given primID and the activation lock (``Activ lock'') which enables/disables changes to ``Activ''.
We define these as scalars or boolean as appropriate and allocate them to the M1/M2 as universal properties, shown in \Cref{tab:slipnet}.
Finally, for simplicity we will use \texttt{primID1} to denote edges and \texttt{primID2} to denote destination vertices.

\begin{table}[!htb]
  \caption{
    Slipnet data under the \textit{CNSM} allocation.
    Items in \textbf{bold} denote data objects used for running Copycat-like activation dynamics. 
    Note, for example, how \textit{destination vertex} is represented as a pointer and stored in array C2 (making it a primID). 
    Similarly, note how the concept of slip lock is represented as a bool and stored within either M1 or M2 arrays.
  }
\centering
\begin{tabular}{llll}
\toprule
Content                              & Type     & Linknode mapping     & Array                  \\
\midrule
\textbf{\textit{source vertex}}      & pointer  & \texttt{head ID}     & N1                     \\
Next linknode                        & address  & \texttt{next}        & N2                     \\
\textbf{\textit{edge label}}         & pointer  & \texttt{primID1}     & C1                     \\
\textbf{\textit{destination vertex}} & pointer  & \texttt{primID2}     & C2                     \\
Subordinate I                        & reserved & \texttt{prop1}       & S1                     \\
Subordinate II                       & reserved & \texttt{prop2}       & S2                     \\
\textbf{Conceptual depth}            & scalar   & \texttt{prop1}       & M1 (\textit{headnode}) \\
\textbf{Activ}                       & scalar   & \texttt{prop1}       & M1 (\textit{headnode}) \\
\textbf{Activ lock}                  & bool     & \texttt{prop1}       & M1 (\textit{headnode}) \\
\textbf{Conductance}                 & scalar   & \texttt{prop1/prop2} & M1/M2 (1/linknode)     \\
\textbf{Slip lock}                   & bool     & \texttt{prop1/prop2} & M1/M2 (1/primID)       \\
\hline
\end{tabular}
\label{tab:slipnet}
\end{table}

With these fields represented, we can give one illustrative activation update.
For an iteration of the activation propagation from the source vertex to the destination vertex, we execute the following pseudocode in the current linknode:

\begin{verbatim}
    if primID1.actiLock == 0:
        primID1.activ = primID1.activ * primID1.conceptualDepth
                        + currLink.head.activ * currLink.conductance
\end{verbatim}

If the edge headnode is not activation-locked, the first term applies decay through conceptual depth ($\leq 1$).
The second adds a fraction ($\geq 0$) of the source headnode activation, controlled by linknode conductance.
Then we continue with:

\begin{verbatim}
    if (primID1.activ > slipnet.threshold) and (primID2.slipLock == 0):
        currLink.head.slippingFrom.append(primID2.head)
\end{verbatim}
which checks whether the activation of the edge headnode (\verb|primID1|) is greater than a preset threshold value, and whether the edge to the destination vertex (via \verb|primID2|) is not slip-locked.
If both conditions are met, the destination vertex is added to the list of slippage candidates (\verb|currLink.head.slippingFrom|) of the source vertex.
We note how assigning slip locks to each individual linknode can selectively activate/deactivate slippage along individual edges in the slipnet's graph.
More complete slippage behavior would require a separate cognitive-model implementation.

The example shows a possible \textit{Views} encoding and update composition for selected slipnet fields.
It is not an implementation or evaluation of cognitive behavior or hardware-supported activation dynamics.
Finally, as a matter of interest, the slipnet of the original Copycat from \citep{mitchell1993analogy} transposed into \textit{Views} format results in 77 headnodes across 11 categories, interconnected by 195 linknodes.

\section{Discussion}

\Cref{tab:model_comparison} positions \textit{Views} relative to \ac{rdf}, \acp{lpg} and earlier recursive graph models \citep{hogan2021knowledge,angles2008survey,pratt1969hierarchical,boley1977directed}.
The comparison distinguishes properties of an abstract data model from common implementation choices: in particular, neither \ac{rdf} nor the \ac{lpg} family prescribes one physical storage organization.
It also separates model-level support from the query languages, optimizers, transactions and reasoning systems that may be built above a graph model.

\begin{table*}[!htb]
  \centering
  \caption{Qualitative comparison of graph database models.}
  \label{tab:model_comparison}
  \small
  \setlength{\tabcolsep}{3pt}
  \begin{tabular}{@{}>{\raggedright\arraybackslash}p{0.13\textwidth}>{\raggedright\arraybackslash}p{0.19\textwidth}>{\raggedright\arraybackslash}p{0.19\textwidth}>{\raggedright\arraybackslash}p{0.19\textwidth}>{\raggedright\arraybackslash}p{0.19\textwidth}@{}}
    \toprule
    Model & \ac{rdf} & \ac{lpg} & Recursive graph models & \textbf{\textit{Views}} \\
    \midrule
    Recursive or relation-local labeling
      & Requires additional representation mechanisms.
      & Relation-local edge properties; nesting depends on the model.
      & Native recursive labels; details vary by formalism.
      & Finite subordinate chains on either primID. \\
      \hline
    Contextual annotation
      & Named graphs group triples at dataset level; contextual semantics are application-defined.
      & Direct edge properties; nested annotation depends on the model.
      & Native where the formalism defines contextual scope.
      & Parent- and side-scoped subordinate chains. \\
      \hline
    Traversal and query support
      & Mature SPARQL graph-pattern and path querying.
      & Mature adjacency traversal and property filtering.
      & Depends on the accompanying language or implementation.
      & Pointer traversal plus AAR/CAR/CAR2; no full query layer. \\
      \hline
    Physical storage organization
      & System dependent.
      & System dependent.
      & No common physical organization.
      & Linked chains in aligned arrays. \\
      \hline
    Hardware mapping
      & Hardware-neutral; accelerator mappings exist.
      & Hardware-neutral; adjacency layouts can be accelerated.
      & Proposal-dependent; not inherent.
      & Explicit ASOCA-oriented associative-array mapping. \\
    \bottomrule
  \end{tabular}
\end{table*}

The table identifies a specific design trade-off rather than a dominance relation.
\textit{Views} makes recursively scoped labels and an associative array mapping explicit.
RDF has mature standards, while both RDF and property-graph ecosystems offer substantially more mature query tooling and operational database functionality.
Mappings among these models remain schema-dependent: for example, edge-list representations can be transformed into adjacency-list representations \citep{arifuzzaman2015fast}, and \textit{Views} main chains have a similar organization.
However, preservation of statement identity and contextual scope must be checked rather than assumed.

Adjacency-list and compressed sparse row (CSR) layouts primarily organize graph topology for neighbor enumeration; physical organization matters when traversal incurs external-memory I/O \citep{munagala1999complexity}.
\textit{Views} spans model and implementation concerns by specifying relation ownership, recursively scoped contextual information and an associative execution mapping.
A projection of main-chain connectivity could in principle use an adjacency-list or CSR layout, but relation labels, subordinate-chain scope and associative field semantics would require aligned auxiliary structures.
Conversely, fixed-width \textit{Views} fields and pointer-linked chains can incur storage and pointer-traversal costs that compact topology-only layouts may avoid.

\begin{figure}[!htb]
  \centering
  \subfloat[]{
    \begin{tikzpicture}[
        scale=0.5, transform shape,
        box/.style={rectangle, draw, minimum width=1.1cm, minimum height=1.1cm, text width=1.1cm, text centered},
        pointer/.style={-stealth, thick}
    ]
    
    \node[box, fill=red!30] (TH) {Tom Hanks};
    \node[box, right=0pt of TH] (ThCdr) {};

    \node[box, right=1.5cm of ThCdr, fill=orange!30] (ActFilm) {\texttt{0x1}};
    \draw[pointer] (ThCdr.east) -- (ActFilm.west);
    \node[box, right=0pt of ActFilm] (ActFilmCdr) {};
    
    \node[box, below=1cm of ActFilm] (ActAs) {};
    \draw[pointer] (ActFilm.south) -- (ActAs.north);
    \node[box, right=0pt of ActAs] (ActAsCdr) {};

    \node[box, right=1.5cm of ActAsCdr] (ThisFilm) {this film};
    \draw[pointer] (ActAsCdr.east) -- (ThisFilm.west);
    \node[box, right=0pt of ThisFilm] (ThisFilmCdr) {};

    \node[right=0.5cm of ThisFilmCdr] (ThisFilmNil) {\texttt{nil}};
    \draw[pointer] (ThisFilmCdr.east) -- (ThisFilmNil.west);

    \node[box, below=1cm of ActAs] (ActIn) {act in};
    \draw[pointer] (ActAs.south) -- (ActIn.north);
    \node[box, right=0pt of ActIn] (ActInCdr) {};

    \node[box, right=1.5cm of ActInCdr, fill=green!30] (AsSully) {\texttt{0xe}};
    \draw[pointer] (ActInCdr.east) -- (AsSully.west);
    \node[box, right=0pt of AsSully] (AsSullyCdr) {};
    \node[right=0.4cm of AsSullyCdr] (AsSullyNil) {\texttt{nil}};
    \draw[pointer] (AsSullyCdr.east) -- (AsSullyNil.west);

    \node[box, below=1cm of AsSully] (As) {as};
    \draw[pointer] (AsSully.south) -- (As.north);
    \node[box, right=0pt of As] (AsCdr) {};

    \node[box, right=1.5cm of AsCdr] (Sully) {Sully};
    \draw[pointer] (AsCdr.east) -- (Sully.west);
    \node[box, right=0pt of Sully] (SullyCdr) {};
    
    \node[right=0.5cm of SullyCdr] (SullyNil) {nil};
    \draw[pointer] (SullyCdr.east) -- (SullyNil.west);

    \node[box, right=6cm of ActFilmCdr, fill=orange!30] (WonOscars) {\texttt{0x2}};
    \draw[pointer] (ActFilmCdr.east) -- (WonOscars.west);
    \node[box, right=0pt of WonOscars] (WonOscarsCdr) {};

    \node[right=0.5cm of WonOscarsCdr] (WonOscarsNil) {\texttt{nil}};
    \draw[pointer] (WonOscarsCdr.east) -- (WonOscarsNil.west);

    \node[box, below=1cm of WonOscars] (Won) {won};
    \draw[pointer] (WonOscars.south) -- (Won.north);
    \node[box, right=0pt of Won] (WonCdr) {};

    \node[box, right=1.5cm of WonCdr] (OscarsFor) {};
    \draw[pointer] (WonCdr.east) -- (OscarsFor.west);
    \node[box, right=0pt of OscarsFor] (OscarsForCdr) {};

    \node[right=0.5cm of OscarsForCdr] (OscarsForNil) {\texttt{nil}};
    \draw[pointer] (OscarsForCdr.east) -- (OscarsForNil.west);

    \node[box, below=1cm of OscarsFor] (Oscars) {2 Os-cars};
    \draw[pointer] (OscarsFor.south) -- (Oscars.north);
    \node[box, right=0pt of Oscars] (OscarsCdr) {};

    \node[box, right=1.5cm of OscarsCdr, fill=green!30] (ForBest) {\texttt{0xf}};
    \draw[pointer] (OscarsCdr.east) -- (ForBest.west);
    \node[box, right=0pt of ForBest] (ForBestCdr) {};

    \node[right=0.5cm of ForBestCdr] (ForBestNil) {nil};
    \draw[pointer] (ForBestCdr.east) -- (ForBestNil.west);

    \node[box, below=1cm of ForBest] (For) {for};
    \draw[pointer] (ForBest.south) -- (For.north);
    \node[box, right=0pt of For] (ForCdr) {};

    \node[box, right=1.5cm of ForCdr] (BestActor) {best actor};
    \draw[pointer] (ForCdr.east) -- (BestActor.west);
    \node[box, right=0pt of BestActor] (BestActorCdr) {};

    \node[right=0.5cm of BestActorCdr] (BestActorNil) {\texttt{nil}};
    \draw[pointer] (BestActorCdr.east) -- (BestActorNil.west);

    \end{tikzpicture}
  }
  \hfil
  \subfloat[]{
    \begin{tikzpicture}[scale=0.9, transform shape]
      \node (TH) [head, minimum width=0] {Tom Hanks};

      \node (ActFilm) [ndot, below of=TH] {\tiny \texttt{0x1}};
      \draw [nextptr] (TH) -- (ActFilm);
      \node (ActIn) [pid1, yshift=0.1cm, xshift=0.5cm, right of=ActFilm] {act in};
      \draw [proplink] (ActFilm) to [bend left=30] (ActIn);
      \node (ThisFilm) [pid2, yshift=0.1cm, xshift=-0.5cm, left of=ActFilm] {this film};
      \draw [proplink] (ActFilm) to [bend right=30] (ThisFilm);

      \node (AsSully) [propdot1, xshift=1.5cm, right of=ActFilm] {\tiny \texttt{0xe}};
      \draw [proplink] (ActFilm) to [bend left=30] (AsSully);
      \node (As) [pid1, right of=AsSully] {as};
      \draw [proplink] (AsSully) to [bend left=15] (As);
      \node (Sully1) [pid2, left of=AsSully] {Sully};
      \draw [proplink] (AsSully) to [bend right=15] (Sully1);


      \node (WonOscars) [ndot, yshift=-0.7cm, below of=ActFilm] {\tiny \texttt{0x2}};
      \draw [nextptr] (ActFilm) -- (WonOscars);
      \node (Won) [pid1, xshift=0.5cm, right of=WonOscars] {won};
      \draw [proplink] (WonOscars) to [bend left=30] (Won);
      \node (2Oscars) [pid2, yshift=0.1cm, xshift=-0.7cm, left of=WonOscars] {2 Oscars};
      \draw [proplink] (WonOscars) to [bend right=30] (2Oscars);


      \node (ForBestActor) [propdot2, xshift=-1.5cm, left of=WonOscars] {\tiny \texttt{0xf}};
      \draw [proplink] (WonOscars) to [bend right=30] (ForBestActor);
      \node (For) [pid1, right of=ForBestActor] {for};
      \draw [proplink] (ForBestActor) to [bend left=15] (For);
      \node (BestActor) [pid2, text width=0.8cm,  left of=ForBestActor] {best actor};
      \draw [proplink] (ForBestActor) to [bend right=15] (BestActor);

      \node (eoc) [next, below of=WonOscars, yshift=-0.5cm, fill=orange!30, minimum width=1cm] {\ac{eoc}};
      \draw [nextptr] (WonOscars) -- (eoc);
    \end{tikzpicture}
    \label{fig:th_twin}
  }
  \caption{
    Structural correspondence between (a) one possible Lisp \texttt{cons} encoding and (b) the \textit{Views} \textit{Tom Hanks} chain expanded from the walkthrough in \Cref{tab:worked_views}.
    Matched addresses and colors identify corresponding list entries.
    Subordinate linknodes encode contextual relationships such as ``2 Oscars for best actor'', while \texttt{nil} and \ac{eoc} terminate their respective lists.
  }
  \label{fig:lisp_example}
\end{figure}

List-based structures provide a complementary way to interpret the chain constructed in \Cref{tab:worked_views}.
Lisp's \texttt{cons}, \texttt{car} and \texttt{cdr} primitives have long supported hierarchical and relational data in symbolic computation \citep{mccarthy1960recursive}.
Under the particular encoding in \Cref{fig:lisp_example}, a selected primary pointer plays a role analogous to \texttt{car}, \texttt{next} plays a role analogous to \texttt{cdr}, and \texttt{NULL} is analogous to \texttt{nil}.
In this framework, a \textit{Views} linknode corresponds to a Lisp \texttt{cons} cell: it pairs a sublist with a \texttt{next} pointer to the subsequent linknode (another \texttt{cons} cell) in a list.

From the perspective of logic and reasoning in \ac{ai}, \textit{Views} should be read as a \ac{gdb} model rather than as a complete reasoning formalism.
Logical, rule-based, graph-based or hybrid procedures could be implemented above the model by selecting schemas and composing retrieval operations.
In this sense, the contribution is complementary to established reasoning systems: \textit{Views} provides a recursively labelable and hardware-aware substrate on which such systems may store, retrieve and traverse structured knowledge.

Beyond the crisp concepts used in the present examples, more advanced forms of graph-based representation and analysis include fuzzy and N-fuzzy graphs with domination and structural measures \citep{rana2026alpha,ali2026structural}, vertex-betweenness centrality \citep{nandi2026vertex}, and complex hesitant-fuzzy decision-making \citep{farooq2026complex}.
Application-specific \textit{Views} schemas could provide a storage and retrieval basis for such extensions by attaching membership information to concepts or relations, including relation-local values through subordinate chains, and by exposing graph topology through traversal and associative retrieval.
The corresponding fuzzy semantics, aggregation operators and graph algorithms would remain higher-level procedures and are not defined or evaluated here.

\subsection{Limitations and Scope}
We further point out some limitations of the current work and clarify its intended scope.
The evidence in this paper has three distinct boundaries: the worked examples demonstrate representation and retrieval rather than complete reasoning; the \acs{ldbc} \acs{snb} study measures storage rather than query execution; and the K-hop study characterizes one host--accelerator workload over a fixed graph image on the \ac{asoca3} \ac{fpga}, rather than \ac{asoca2}, ASIC or general graph-processing performance.

\textit{Views} is not a replacement for \ac{rdf}, \acp{lpg}, SPARQL, Cypher or existing \ac{gdb} engines.
No standards-compliant query processor, complete transaction system, end-to-end multi-hop question-answering workload, \ac{rag} pipeline or neuro-symbolic application is implemented or evaluated here.
The Copycat-inspired example does not validate a cognitive model.
The K-hop study retains its fixed source cohort, hop-level result cardinalities and descriptive aggregates, but not the raw latency observations, exact timer endpoints or complete graph-loading details.
Commercial confidentiality also limits the \ac{asoca3} description to benchmark-relevant top-level configuration and public primitive semantics.
Future work should extend hardware prototype evaluation to larger resident graphs, repeated latency and throughput measurements, workload power and energy, controlled CPU/GPU and contemporary graph-accelerator baselines, and integration with end-to-end query and advanced reasoning systems under independently reproducible conditions.

\section{Conclusions}
In this paper, we have proposed a \ac{gdb} model named \textit{Views}, which supports recursively labeled graphs and organizes their representation as linknode chains mapped to an associative execution model.
We have characterized analytical and materialized storage footprint for \acs{ldbc} \acs{snb} initial snapshots at three scales, separating schema choice, entry width, dictionaries and native database-store boundaries, and reported directed, up-to-$k$ traversal characteristics for a resident road-network graph on the later \ac{asoca3} \ac{fpga}.
The storage results demonstrate allocation and width specialization rather than an intrinsic storage advantage for \textit{Views}, while the K-hop results show increasing mean returned-vertex count and mean latency with hop bound over the fixed graph image, with accelerator-core and host--accelerator system timings kept distinct.
The semantic-inference and Copycat-inspired cases illustrate representation and primitive composition rather than application-level reasoning performance or cognitive validation.
Taken together, these results position \textit{Views} as a model--architecture co-design for associative storage and selected graph traversal under the stated conditions, rather than as a replacement for established graph models and database systems.
Complete database functionality, general reasoning acceleration and end-to-end AI applications remain unevaluated.

\section*{Conflict of Interest Statement}
The authors declare that the research was conducted in the absence of any commercial or financial relationships that could be construed as a potential conflict of interest.

\section*{Author Contributions}
YY, AW, YP and AS contributed to the conception and development of the work.
YP, YY, AW and AS contributed to the hardware design and implementation.
YY prepared the literature review and the initial manuscript draft, and contributed to data curation and visualization.
YY and AS contributed to manuscript revision.
AS contributed to project supervision and administration.
All authors contributed to manuscript finalization.

\section*{Funding}
This study was supported by the Engineering and Physical Sciences Research Council (EPSRC) under Grant EP/V008242/2 and Scottish Enterprise High Growth Spinout Programme (HGSP) \textit{Atlas}, funding reference number PS7305129O.

\section*{Data Availability Statement}
The storage-benchmark metadata supporting the \acs{ldbc} study are included as arXiv ancillary files in \path{anc/storage-benchmark/}.
The K-hop benchmark results are reported in aggregate in the article; implementation details concern commercially confidential hardware and lie outside the scope of this work.

\section*{Generative AI Statement}
Generative AI technologies were used to assist with language editing. The authors reviewed and edited the output and take full responsibility for the content of this manuscript.

\vskip2pc

\bibliographystyle{Frontiers-Harvard}

\bibliography{references}

\end{document}